%% file: main.tex
\documentclass[final,3p,times]{elsarticle}

\usepackage{amssymb}
\biboptions{comma,sort&compress}
\usepackage{amsthm}
\usepackage{amsmath}

\usepackage{verbatim}

\usepackage{graphicx}
\usepackage{subcaption}
\usepackage{adjustbox}
\usepackage{algorithm}
\usepackage{algpseudocode}

\usepackage{xcolor}

\usepackage[acronym]{glossaries}
\loadglsentries[\acronymtype]{acro}

\usepackage{hyperref}

\begin{document}

\newpage
\begin{frontmatter}



\title{\textit{a priori} Uncertainty Quantification of Reacting Turbulence Closure Models using Bayesian Neural Networks}


\author[UT]{Graham Pash\corref{cor1}}
\ead{gtpash@utexas.edu}

\affiliation[UT]{organization={Oden Institute for Computational Engineering and Sciences},
            addressline={The University of Texas at Austin, 201 E. 24th Street}, 
            city={Austin}, 
            state={TX},
            postal={78712},
            country={USA}}

\author[NREL]{Malik Hassanaly}
\author[NREL]{Shashank Yellapantula}

\affiliation[NREL]{
            organization={Computational Science Center, National Renewable Energy Laboratory, 15013 Denver West Parkway}, 
            city={Golden},
            state={CO},
            postal={80401},
            country={USA}}

\input{abstract}


\end{frontmatter}



\input{intro}
\input{formulation}
\input{results}

\input{discussion}
\input{conclusion}
\input{credit}
\input{acknowledgements}


\bibliographystyle{elsarticle-num} 
\bibliography{refs}






\newpage
\input{appendix}


\end{document}

%% file: acro.tex
\newacronym{LES}{LES}{large eddy simulation}
\newacronym{SFS}{SFS}{sub-filter scale}
\newacronym[plural=DNSs]{DNS}{DNS}{direct numerical simulations}
\newacronym[plural=QoIs,firstplural=quantities of interest (QOIs)]{QoI}{QoI}{quantity of interest}
\newacronym[plural=BNNs,firstplural=Bayesian neural networks (BNNs)]{BNN}{BNN}{Bayesian neural network}
\newacronym[plural=DNNs]{DNN}{DNN}{deterministic neural network}
\newacronym{LRM}{LRM}{linear relaxation model}
\newacronym{ELBO}{ELBO}{evidence lower bound}
\newacronym{KL}{KL}{Kullback-Liebler}
\newacronym{OOD}{OOD}{out-of-distribution}
\newacronym{SBO}{SBO}{soft Brownian offset}
\newacronym{NF}{NF}{normalizing flow}
\newacronym{OED}{OED}{optimal experimental design}
\newacronym{RANS}{RANS}{Reynolds-averaged Navier-Stokes}
\newacronym{ML}{ML}{machine learning}
\newacronym{SciML}{SciML}{scientific machine learning}
\newacronym{MLP}{MLP}{multi-layer perceptron}
\newacronym[plural=PINNs,firstplural=physics-informed neural networks (PINNs)]{PINN}{PINN}{Physics Informed Neural Network}

%% file: abstract.tex
\begin{abstract}
While many physics-based closure model forms have been posited for the \gls{SFS} in \gls{LES}, vast amounts of data available from \gls{DNS} create opportunities to leverage data-driven modeling techniques. Albeit flexible, data-driven models still depend on the dataset and the functional form of the model chosen. Increased adoption of such models requires reliable uncertainty estimates both in the data-informed and out-of-distribution regimes. In this work, we employ \glspl{BNN} to capture both epistemic and aleatoric uncertainties in a reacting flow model. In particular, we model the filtered progress variable scalar dissipation rate which plays a key role in the dynamics of turbulent premixed flames. We demonstrate that \gls{BNN} models can provide unique insights about the structure of uncertainty of the data-driven closure models. We also propose a method for the incorporation of out-of-distribution information in a \gls{BNN}, which can be used for out-of-distribution query detection. The efficacy of the model is demonstrated by \textit{a priori} evaluation on a dataset consisting of a variety of flame conditions and fuels.
\end{abstract}

%% file: intro.tex
\section{Introduction}
\label{sec:intro}

Numerical simulations of turbulent reacting flows are usually associated with high computational cost due to the large range of spatiotemporal scales that need to be resolved \cite{pope2000turbulent}. Concurrent with significant computational advances \cite{alexander2020exascale}, several modeling approaches have since been devised to avoid resolving the smallest spatio-temporal scales. The two most prominent are \gls{RANS} \cite{alfonsi2009rans} and \gls{LES} \cite{pitsch2006large}. We will focus on \gls{LES} in this work, but the approximations made in both approaches introduce the need for closure modeling \cite{echekki2010turbulent,poinsot2005theoretical}. In \gls{LES}, one only resolves the largest fluid flow scales. This is achieved by applying a low-pass filter to the original governing equations, which in turn requires to model the sub-filter scales. The modeling of this closure term is the primary focus of this paper. While several physics-based modeling strategies for closure terms have emerged, they tend to make strict assumptions about the unresolved scales \cite{fiorina2005premixed,sagaut2005large}. While physics models increasingly avoid these assumptions \cite{akram2022approximate}, data-driven strategies are becoming ever more popular, given their flexibility \cite{ihme2022combustion}, data availability, and the development of \gls{SciML} frameworks \cite{karniadakis2021physics,kovachki2023neural,peherstorfer2016data,duraisamy2019turbulence,duraisamy2021perspectives,o2024derivative}. Recent \gls{SciML} developments have focused on enforcing known physics constraints, such as with \glspl{PINN} \cite{raissi2019physics,cai2021physics}. This approach can be useful if a deconvolution procedure is used to compute closure terms \cite{bode2021using,hassanaly2022adversarial}. However, when directly approximating the closure term, no physics-law are typically available, and a purely data-driven approach is more appropriate.

There exists a considerable body of literature applying data-driven techniques for closure modeling broadly \cite{sanderse2024scientific}. For turbulent combustion applications, multiple data-driven strategies have been shown to be accurate approximators of closure terms, such as artificial neural networks \cite{vollant2017subgrid,barwey2021data,maulik2019subgrid,raman2019emerging,van2024energy,ihme2022combustion}, convolutional neural networks \cite{lapeyre2019training, nikolaou2019progress}, and neural ordinary differential equations \cite{kang2023learning}. While flexible, data-driven closure modeling strategies still depend on the model form and the training procedure, both of which are typically optimized via a hyper-parameter search. In addition, the choice of the dataset itself can induce significant variability on the closure model \cite{hassanaly2023uniform}. These choices can result in a significant uncertainty thereby motivating the need to equip these \gls{ML} methods with reliable uncertainty estimates that can be propagated through the governing equations. From a practical perspective, providing objective uncertainty estimates of data-driven models is necessary before they can be confidently adopted or not by the engineering community. Uncertainty estimates are also needed for decision-making tasks, such as design optimization, where one relies on numerical simulations to optimize specific quantities of interest, e.g. ignition time \cite{tang2021probabilistic, jaravel2019numerical}, maximum temperature \cite{masquelet2017uncertainty}, or mechanical structure \cite{abdelsalam2023comparative,abdelsalam2024optimizing}.

A variety of machine learning methods have been developed to model uncertainty~\cite{hullermeier2021aleatoric,abdar2021review,nelsen2021random}. Popular methods to quantify uncertainty in neural networks include the dropout method \cite{srivastava2014dropout, jospin2022hands} and its Bayesian interpretation \cite{gal2016dropout}. While inexpensive and easy to implement, dropout generates probabilistic predictions at inference time but does not quantify the model parameter uncertainty \cite{anderson2020meaningful}. In computational science and engineering, Gaussian processes have long been applied to uncertainty quantification tasks \cite{rasmussen2006gaussian} owing to their flexibility and interpretability. The primary drawback of Gaussian processes is the $\mathcal{O}(n^3)$ computation expense to train and $\mathcal{O}(n^2)$ cost to evaluate due to the required matrix inversion and multiplication, where $n$ is the data size. Despite efforts, scalability for large datasets remains a challenge \cite{snelson2008flexible,liu2020gaussian,banerjee2013efficient,wang2016bayesian,bauer2016understanding}. New approaches are needed to handle massive datasets generated by \gls{DNS}, especially in the context of rapid online evaluation within \gls{LES} codes.

\glspl{BNN} are an attractive method to estimate and predict modeling uncertainty due to their ability to ingest large amounts of data, relatively fast inference cost (compared to Gaussian processes), rigorous characterization of uncertainty, and expressivity. Recent advancements employing variational inference \cite{jospin2022hands} have made the training of \glspl{BNN} tractable for large models and amenable to large datasets \cite{graves2011practical,blundell2015weight}. \glspl{BNN} reformulate deterministic deep learning models as point-estimators and emulate the construction of an ensemble of neural nets by assigning a probability distribution to each network parameter \cite{mackay1995probable}. Thus, they generate a predictive distribution by sampling the parameter distributions and collecting the resulting distribution of point estimates. 

\subsection{Contributions}
\label{sec:intro-contribution}
The central contribution of this paper is the development and application of a Bayesian neural network model to modeling the sub-filter progress variable dissipation rate of premixed turbulent flames \cite{yellapantula2021deep}. In turbulent combustion modeling, uncertainty quantification tasks primarily target kinetic rates \cite{braman2013bayesian,mueller2013chemical,wang2015combustion,najm2009uncertaintyChem}, operating and boundary conditions \cite{masquelet2017uncertainty} and model coefficient in closure models \cite{khalil2015uncertainty}, but do not address uncertainties of data-driven closure models. Recent applications of \glspl{BNN} for combustion modeling have mostly focused on the prediction of macroscopic quantities such as ignition delay \cite{mccartney2022reducing}, fuel properties \cite{oh2024learning}, or to accelerate data assimilation \cite{croci2021data}. To our knowledge, this work is the first to explore the use of \glspl{BNN} for quantifying both \textit{epistemic} and \textit{aleatoric} uncertainties in data-driven closure models \cite{najm2009uncertainty}. While the framework developed and demonstrated in the context of progress variable dissipation rate, the techniques may be readily applied to other data-driven closure models.

Beyond the development of the model itself, we focus on two important applications. First, we utilize the learned epistemic uncertainty to derive physical insights about the potential weaknesses of the model. This knowledge can be used, for example, to inform what future data collection should be performed to improve the model's predictive quality. Second, we provide a method for endowing a \gls{BNN} model with meaningful predictive power in the out-of-distribution setting, i.e. when tasked with extrapolation in a regression setting. This allows one to strongly enforce a prior or detect scenarios where the data-driven model may fail and resort to another method or heuristic.

Like other data-driven closure models, \glspl{BNN} can be introduced into existing reacting \gls{LES} code in place of physics-based closure models, allowing for closure modeling prediction with uncertainty quantification. The integration of the \gls{BNN} within a reacting flow solver will be the object of future work. The remainder of this paper is organized as follows. In Section~\ref{sec:formulation}, we formulate the target problem, including sources of uncertainty, dataset generation, and Bayesian neural network modeling. Numerical experiments and their results assessing the model performance are presented in Section~\ref{sec:results}. We discuss the utility of the model for downstream applications and its limitations in Section~\ref{sec:discussion}. Concluding remarks are given in Section~\ref{sec:conclusion}.

%% file: formulation.tex
\section{Formulation}
\label{sec:formulation}
In this section we first review the nature of aleatoric and epistemic uncertainty in Section~\ref{sec:uncertainty}. Next, we discuss generation of the training data in Section~\ref{sec:data}. We then provide an overview of Bayesian neural networks in Section~\ref{sec:modeling-bnn} as well as practical considerations (Section~\ref{sec:modeling-prior} and Section~\ref{sec:modeling-extrapolation}). We conclude by providing the architecture for our experiments in Section~\ref{sec:modeling-specification}.

\subsection{Sources of Uncertainty}
\label{sec:uncertainty}
While there are multiple sources of uncertainty arising from model inputs, numerical errors, or experiments, in this work we primarily consider the uncertainties driven by the data that the model trains on. In the machine learning context, the use of deep neural networks is typically justified by a universal approximation theorem \cite{hornik1989multilayer} valid in the asymptotic limit of data availability. Thus, we concern ourselves with uncertainty in the learned parameters and resultant functional representation due to the empirical data distribution used for training. 

\begin{figure}[!hbt]
    \centering
    \includegraphics[width=0.65\textwidth]{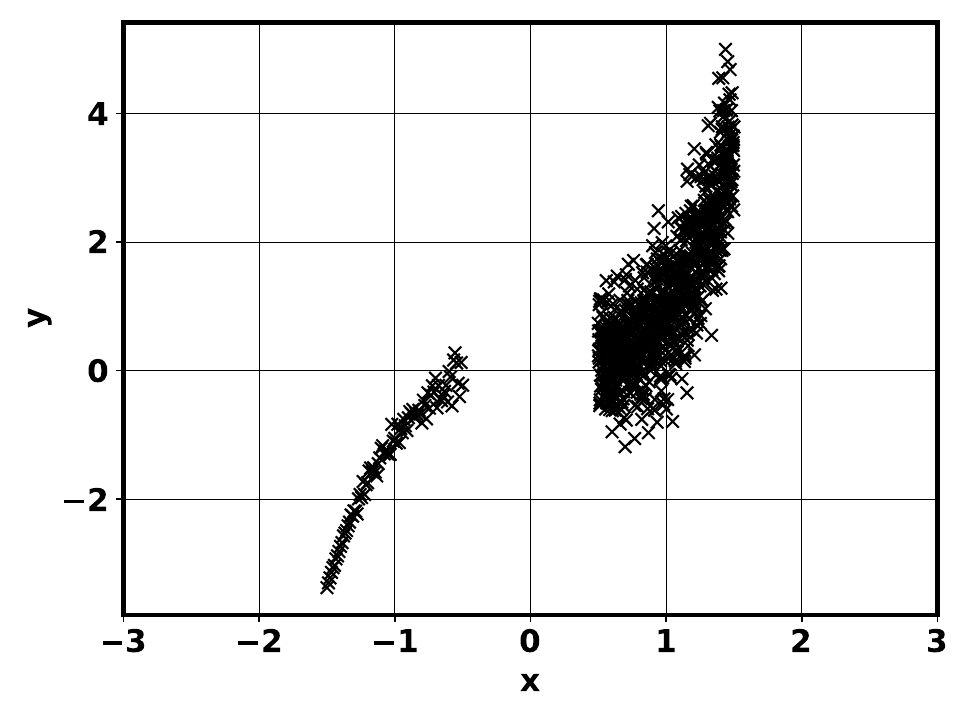}
    \caption{A dataset that has a region of high epistemic uncertainty and low aleatoric uncertainty (left of origin) as well as a region with low epistemic uncertainty and high aleatoric uncertainty (right of origin).}
    \label{fig:uncertainty-cartoon}
\end{figure}

Broadly speaking, uncertainties can be categorized as either aleatoric or epistemic \cite{smith2013uncertainty,der2009aleatory,najm2009uncertainty}. As an illustrative example, consider the example shown in Fig.~\ref{fig:uncertainty-cartoon}. The underlying generating function for the data is 
\begin{equation}
    y = x ^3 + 0.1 (1.5+x)\varepsilon,
\end{equation}
where $\varepsilon\sim\mathcal{N}(0,\sigma^2)$. Here, the noise in the data increases as the variable $x$ increases, however only 50 data points are retained in the region on the left, compared to 500 data points in the region on the right. The left region has low aleatoric uncertainty due to the lack of noise but has high epistemic uncertainty due to the lack of data. On the other hand, any model on the right would be well informed by the quantity of the data and so would have low epistemic uncertainty, but would have a relatively high level of aleatoric uncertainty due to high noise level.

In the context of data-driven closure modeling, epistemic uncertainties arise from a lack of knowledge when too few training data are available. An ideal \gls{LES} model in the sense of Langford et al.~\cite{langford1999optimal} obtained via a data-driven method, could be subject to epistemic uncertainty when there does not exist sufficient data to confidently determine the optimal estimator of the closure term. Epistemic uncertainty also presents itself in the extrapolation setting, which may be viewed as an extreme case of lacking data availability. 

Aleatoric uncertainty is also known as statistical or stochastic uncertainty and is inherent to a problem or experiment. It is irreducible in the sense that it cannot be reduced by additional data collection. A clear example is the measurement tolerance of a sensor: additional data collection would not inform one to a greater precision than the instrument rating. In the data-driven closure modeling context, aleatoric uncertainty primarily presents itself as coarse–graining or filtering uncertainty. The act of filtering destroys information, as there are potentially many \gls{DNS} realizations that, when filtered, would present identical filtered measurements. Inferring a mapping from filtered to unfiltered data is therefore subject to an irreducible uncertainty \cite{langford1999optimal,adrian1990stochastic,pope2010self, hassanaly2022adversarial}. Likewise, the selection of input features can also be seen as a coarse-graining step.

From a practical perspective in the context of closure modeling, a quantified epistemic uncertainty allows one to decide where to collect additional data samples before retraining a model. The aleatoric uncertainty may help reassess the choice of input features. Quantifying epistemic and aleatoric uncertainties is the first step to propagate them through a fluid simulation, and incorporate their effect into decision-making tasks.

\subsection{Reacting Turbulent Flow Dataset Generation}
\label{sec:data}
The dataset used in this work is obtained from several \gls{DNS} of turbulent premixed flames \cite{lapointe2015differential,savard2017effects}. The reader is referred to Yellapantula et al.~\cite{yellapantula2021deep} for a more detailed description of the data generation and preparation processes. Hereafter, only key details of the dataset are repeated. The \gls{DNS} data was Favre filtered using Gaussian filter kernels with different filter sizes to prepare input features for \textit{a priori} model evaluation that would match the \gls{LES} features available in the \textit{a posteriori} setting. The objective is to learn an optimal mapping between the filtered quantities and the quantity of interest, here the unresolved contribution to the filtered progress variable dissipation rate, $\chi_{\rm SFS}$ defined as 
\begin{equation}
     \chi_{\rm SFS} = \widetilde{\chi} - \chi_{\widetilde{C}},
\end{equation}
where $\chi_{\widetilde{C}}$ is the resolved progress variable dissipation rate defined as 
\begin{equation}
     \chi_{\widetilde{C}} = 2\widetilde{D}_C |\nabla \widetilde{C} |^2,
\end{equation}
and $\widetilde{\chi}$ is the total progress variable dissipation rate defined as 
\begin{equation}
     \widetilde{\chi} = \widetilde{2 D_C |\nabla C |^2},
\end{equation}
where $D_C$ is the progress variable diffusivity, $C$ is the progress variable, and $\widetilde{.}$ denotes the Favre filtering operation.

The relevance of approximating closure models for the subfilter progress variable dissipation rate has been extensively documented elsewhere and is not repeated here \cite{gao2014scalar,yellapantula2021deep}. The input features are the same as those considered in ~\cite{yellapantula2021deep} and are summarized in Table~\ref{tab:inputs}. The input features were selected based on a feature importance metric \cite{sundararajan2017axiomatic} and only contain filtered quantities as the model is envisioned to be deployed online within a larger \gls{LES} model. Unlike ~\cite{yellapantula2021deep}, the output layer only predicts the unresolved contribution of the progress variable dissipation rate $\chi_{\rm SFS}$. Other derived quantities include the principle rates of filtered strain rate. The principal rates of strain, $\alpha > \beta > \gamma$ are computed using the strain rate tensor constructed from the filtered velocity field. Additionally, the alignments between the eigenvectors and local gradient of the filtered progress variable were added to the list of input features \cite{yellapantula2021deep}. The input space $\mathbf{x}\in\mathbb{R}^{d_i}$ is mapped to an output $\mathbf{y}\in\mathbb{R}^{d_o}$, where $d_i=10$ is the dimensionality of the input space and $d_o$ is the dimensionality of the output space, in this case $d_o=1$. The effect of the number of input parameters on the estimated uncertainties is further discussed in \ref{sec:appendix-12D}.

\begin{table}[!htb]
    \centering
    \begin{tabular}{c|l}
        Input Feature & Description \\
        \hline \\
         $\widetilde{C}$ & Filtered progress variable \\
         $\widetilde{C^{\prime\prime 2}}$ & Filtered progress variable variance \\
         $2\widetilde{D}_C |\nabla \widetilde{C} |^2$ & Resolved progress variable dissipation rate \\
         $\widetilde{D}_{C}$ & Filtered progress variable diffusivity \\
         $\alpha$, $\beta$, $\gamma$ & First, second, and third principal rate of strain \\
         $e_{\alpha,\beta,\gamma} \cdot \nabla \widetilde{C} / |\nabla \widetilde{C} |$ & Alignment of local progress variable gradient with principal eigenvectors
    \end{tabular}
    \caption{Training input features.}
    \label{tab:inputs}
\end{table}

After the selection of the input features, a near-uniform in phase-space sampling \cite{hassanaly2023uniform} was performed by clustering the data with 40 clusters in the input feature space and uniformly selecting data from each cluster. This process allows for appropriately capturing the reaction zone of all the \gls{DNS} cases in the dataset with a reduced number of data points. The final dataset contains $7.88\times 10^6$ data points for training and $2.63\times 10^6$ data points for testing, which is about $10^3$ times less than the total number of \gls{DNS} data points initially available.

\subsection{Bayesian Neural Network Modeling}
\label{sec:modeling-bnn}
Classical neural networks are only capable of deterministically generating point estimates and thus are not well suited for the task of assessing uncertainty. On the other hand, Bayesian neural networks model epistemic uncertainty estimation by placing a parametric distribution over the neural network parameters \cite{denker1990transforming,mackay1992practical,neal2012bayesian,blundell2015weight}. By sampling the parameter distributions, one selects an ensemble of \glspl{DNN}. Note that sampling multiple candidate neural networks to estimate uncertainty is consistent with peer-modeling approaches for uncertainty propagation \cite{mueller2018model}. In particular, we consider the parametric distribution for the weights to be a Gaussian  \cite{blundell2015weight}. The mean and variance of the weights become the parameters to be learned in the \gls{BNN} representation. To delineate between the sampled parameters defining a \gls{DNN} and the distribution placed over the weights in the \gls{BNN} representation, we will set the notation that ``weights'' $\mathbf{w}$ are sampled from a (parametric) distribution placed upon the weights $q(\mathbf{w}|\theta)$, where $\theta$ are the parameters of the \gls{BNN}. Under this notation, $\mathbf{w}\sim q(\mathbf{w}|\theta)$.

In the variational inference formulation of \gls{BNN} training \cite{magris2023bayesian}, one seeks to minimize the \gls{KL} divergence between the weight distribution and the true Bayesian posterior conditioned on the dataset $\boldsymbol{D}$,
\begin{equation}
\theta^* =  \operatorname*{arg\,min}_\theta \operatorname{KL}\left[ q(\mathbf{w}|\theta) || p(\mathbf{w}|\boldsymbol{D}) \right].
\end{equation}
Expansion and rearrangement of this formulation yields the \gls{ELBO} objective function \cite{blundell2015weight},
\begin{equation}
\label{eq:elbo}
\operatorname*{arg\,min}_\theta \underbrace{\operatorname{KL}\left[ q(\mathbf{w}|\theta) || p(\mathbf{w}) \right]}_{\text{prior-informed}} - \underbrace{\mathbb{E}_{q(\mathbf{w}|\theta)}\left[ \log p(\boldsymbol{D}|\mathbf{w}) \right]}_{\text{data-informed}}.
\end{equation}
The first term in the \gls{ELBO} is the \gls{KL} divergence between the learned distribution on the \gls{BNN} parameters and the prior. The second term in the \gls{ELBO} is a data misfit term given by the expected negative log-likelihood of the data over the distribution of plausible models encoded by the distribution of the weights. The consequences of this optimization formulation are explored in the following sections. Hereafter, this probabilistic representation of the model predictions is discussed for regression tasks.

Although the primary interest of modelers that deploy closure models in a reacting flow solver is the epistemic uncertainty, one may wish to evaluate the aleatoric uncertainty as well (see Sec.~\ref{sec:formulation}). One method for doing so is to further parameterize each output dimension so as to capture heteroskedastic uncertainty in the data \cite{kendall2017uncertainties, hoffmann2021deep, nemani2023uncertainty}. We employ the modeling choice from Kendall~and~Gal~\cite{kendall2017uncertainties} of a Gaussian form for the output random variable.

\begin{figure}[!htb]
    \centering
        \begin{subfigure}[b]{0.49\textwidth}
        \centering
        \includegraphics[width=\textwidth]{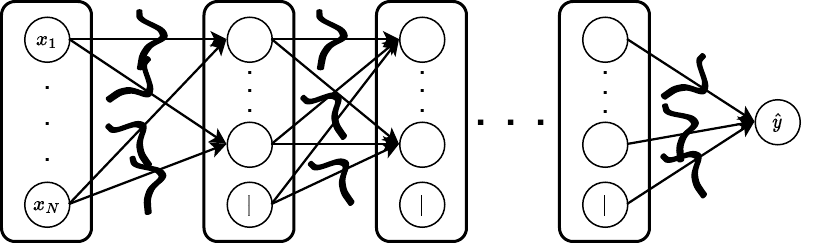}
        \caption{}
        \label{fig:epi-bnn}
    \end{subfigure}
    \\ \vspace{1em}
    \begin{subfigure}[b]{0.49\textwidth}  
        \centering 
        \includegraphics[width=\textwidth]{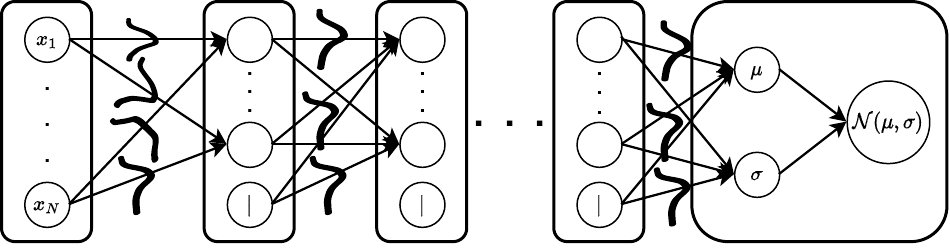}
        \caption{}
        \label{fig:full-bnn}
    \end{subfigure}
    \caption{(a) A \gls{BNN} in the style of \cite{blundell2015weight} that captures epistemic uncertainty only and (b) a \gls{BNN} in the style of \cite{kendall2017uncertainties} that captures both epistemic and aleatoric uncertainty.}
    \label{fig:bnn-architecture}
\end{figure}

In this work, a dense feed-forward neural network parameterization of the closure term is adopted. That is, layers comprise a weight matrix $W$ acting upon the previous layers' outputs, a bias term $b$, and a non-linear activation $\sigma$,
\begin{equation}
    x_{\ell+1} = \sigma_\ell \left(W_\ell x_\ell + b_\ell\right).
\end{equation}
Here $x_\ell$ denotes the output from the last layer, and $\mathbf{x}\in\mathbb{R}^{d_i}$ again represents an input data point. The final layer is mapped to $(\mathbf{y}_{\boldsymbol{\mu}}, \mathbf{y}_{\boldsymbol{\sigma}})^T\in\mathbb{R}^{2d_o}$ in the case of a Kendall~and~Gal~\cite{kendall2017uncertainties} style architecture. Here $\mathbf{y}_{\boldsymbol{\mu}}, \mathbf{y}_{\boldsymbol{\sigma}}$ parameterize the distribution of the output variable as $\mathbf{y}\in\mathbb{R}^{d_o}\sim\mathcal{N}\left(\mathbf{y}_{\boldsymbol{\mu}}, \operatorname{diag}(\mathbf{y}_{\boldsymbol{\sigma}})\right)$. This parameterization allow to model epistemic uncertainty (through the variance of $\mathbf{y}_{\boldsymbol{\mu}}$) the aleatoric uncertainty (through the variance of $\mathbf{y}_{\boldsymbol{\sigma}}$). The difference in the epistemic only architecture and the epistemic and aleatoric architectures is shown graphically for $d_o=1$ in Fig.~\ref{fig:epi-bnn} and Fig.~\ref{fig:full-bnn}, respectively. In this work, we adopt the notation that the output $\mathbf{y}$ is the result of sampling the output of a \gls*{BNN} parameterized by $\theta$ at a datum $\mathbf{x}$ after appropriate sampling of the weights $\mathbf{w}$ and the resultant random variable.

\begin{figure}[!hbt]
    \centering
        \begin{subfigure}[b]{0.49\textwidth}
        \centering
        \includegraphics[width=\textwidth]{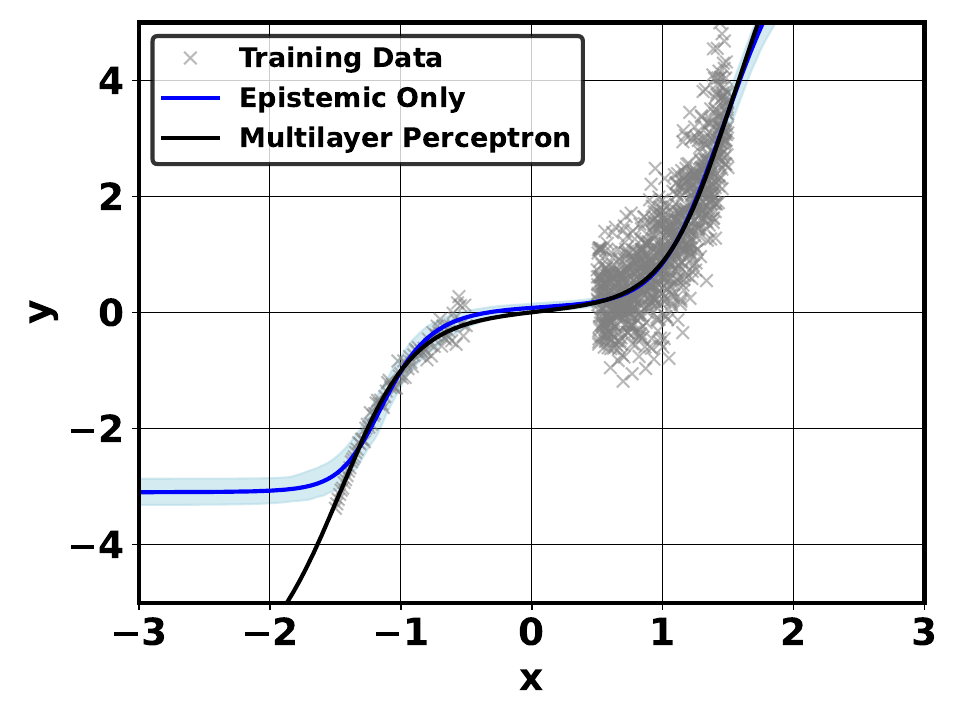}
        \caption{}
        \label{fig:epi-bnn-toy}
    \end{subfigure}
    \hfill
    \begin{subfigure}[b]{0.49\textwidth}  
        \centering 
        \includegraphics[width=\textwidth]{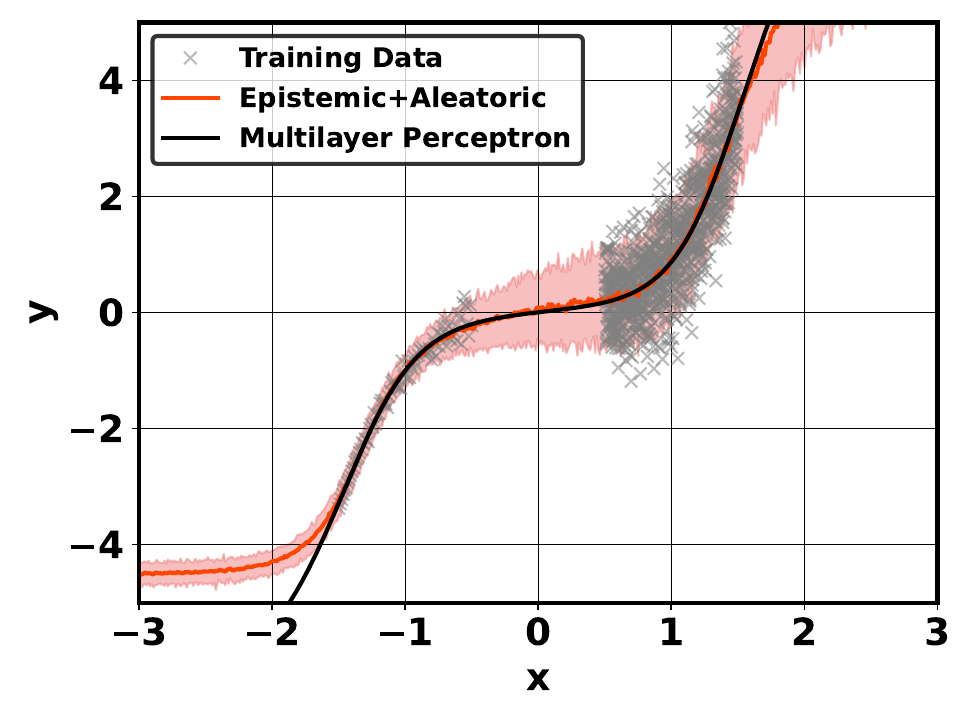}
        \caption{}
        \label{fig:full-bnn-toy}
    \end{subfigure}
    \caption{Comparison of two Bayesian neural networks trained on the dataset presented in Fig.~\ref{fig:uncertainty-cartoon}, (a) models only the epistemic (model-form) uncertainty, while (b) models both epistemic and aleatoric uncertainty.}
    \label{fig:uncertainty-comp}
\end{figure}

In this framework, one can use the law of total variance \cite{saltelli2010variance} to decompose the variance (uncertainty) in the predictions into its epistemic and aleatoric components \cite{chai2018uncertainty,nemani2023uncertainty},
\begin{equation}
    \label{eqn:uncertainty-decomposition}
    \operatorname{Var}(\mathbf{y}) = \underbrace{\mathbb{E}_{q(\mathbf{w}|\theta)} \left[\operatorname{Var}(\mathbf{y} \vert \mathbf{x})\right]}_{\text{aleatoric}} + \underbrace{\operatorname{Var}_{q(\mathbf{w}|\theta)}\left(\mathbb{E}\left[\mathbf{y}\vert \mathbf{x}\right]\right)}_{\text{epistemic}}.
\end{equation}
The predictive variance  $\operatorname{Var}(\mathbf{y})$ is decomposed into an aleatoric component $\mathbb{E}_{q(\mathbf{w}|\theta)} \left[\operatorname{Var}(\mathbf{y} \vert \mathbf{x})\right]$, the mean variability in the estimates, and an epistemic component $\operatorname{Var}_{q(\mathbf{w}|\theta)}(\mathbb{E}[\mathbf{y}\vert \mathbf{x}])$, the variability in the model's mean predictions. For a given $\mathbf{w}\sim q(\mathbf{w}|\theta)$, the predictive mean is given by the first network output, $\mathbf{y}_{\boldsymbol{\mu}}\in\mathbb{R}^{d_o}$. Detailed algorithms for computation of the uncertainties are provided in \ref{sec:appendix-algo}. For a concrete example, the magnitudes of the uncertainties for the one-dimensional example (Fig.~\ref{fig:uncertainty-cartoon}) are shown in Fig.~\ref{fig:uncertainty-comp}. It can be observed that the regions with high and low epistemic uncertainty are appropriately characterized in Fig.~\ref{fig:epi-bnn-toy} and that the predictive envelope captures the entire data distribution in Fig.~\ref{fig:full-bnn-toy}.

\subsubsection{Specification of the Prior}
\label{sec:modeling-prior}
Bayesian inference provides a way to integrate empirical data with prior information about phenomenon being modeled. In practice, the prior drives the first term in the \gls{ELBO} objective function (Eq.~\ref{eq:elbo}), and thus the selection of the prior has a direct impact on the quality of the resultant model \cite{izmailov2021bayesian}. In extreme cases, prior misspecification can only be overcome in the asymptotic limit of data, with the resultant model strictly adhering to the prior belief. 

There are two ways of viewing \gls{BNN} priors: the weight-space view and the functional-space view. Since \gls{BNN}s operate on a parameterized distribution of the weights, $q(\mathbf{w}\vert\theta)$, a prior is inherently defined on the \gls{BNN} parameters. However, the effect of a single weight on the output is difficult to gauge due to interactions with other weights and nonlinear layers. In other words, the lack of interpretability of the neural network parameter prevents one from meaningfully assigning a prior to the weights. In this case, using a non-informative prior such as an isotropic Gaussian could be envisioned, however, this has been shown to be sub-optimal \cite{fortuin2021bayesian}. Often one does have intuition about the outputs and many methods have been derived to ``tune'' priors empirically. These include warm-start methods \cite{ash2020warm}, initial fitting of the model to Gaussian process realizations \cite{flam2017mapping}, and initialization from a trained deterministic model \cite{krishnan2020specifying}. One may be tempted to instead ignore the contribution of the prior altogether, however, it has been shown that direct minimization of the negative log-likelihood is inappropriate for \gls{BNN}s \cite{wei2022performance}. In this work, we employ a ``trainable'' prior that updates the prior mean every epoch while while keeping the variance fixed. This choice changes the interpretation of the \gls{KL} term to that of a data-driven regularization and has connections to the \gls{KL} annealing method in \cite{bowman2015generating}.

\subsubsection{Enforcing Extrapolation Behavior}
\label{sec:modeling-extrapolation}
The aforementioned strategies provide a way to quantify epistemic and aleatoric uncertainty where data is available. Deploying data-driven models in place of physics-based models requires that the phase space spanned by the training data appropriately encompasses the phase space over which the model is queried. In general, this requirement is difficult to guarantee, which has traditionally led to adaptive methods \cite{pope1997computationally} or data-driven methods complemented with synthetic data \cite{chatzopoulos2013chemistry}. In the absence of training data, one would like for extrapolatory predictions to closely match those coming from physics models. Solely relying on the extrapolatory nature of neural network models is unrealistic, especially since it is still not fully understood how neural networks extrapolate \cite{xu2020neural,zhou2022domain}. In the present case, if training data is not available, one may prefer that the sub-filter progress variable dissipation rate should fall back to the predictions of a physics-based model. One way to ensure this behavior would be to detect that a model is called outside of the range of the training data, in which case a separate closure model may be used. However, detecting whether or not an individual input is \gls{OOD} is a challenging problem. An alternative approach is to include data obtained from the physics-based model in the training dataset. To ensure that the synthetic data does not ``pollute'' the original dataset, the data is generated only in the \gls{OOD} region feature space that is separate from the true data distribution, $\boldsymbol{D}$. The synthetic data can be labeled with the prediction of the physics model and an additive noise in case the physics model is subject to uncertainty.

\begin{figure*}[!htb]
    \centering
    \begin{subfigure}[b]{0.49\textwidth}
        \centering
        \includegraphics[width=\textwidth]{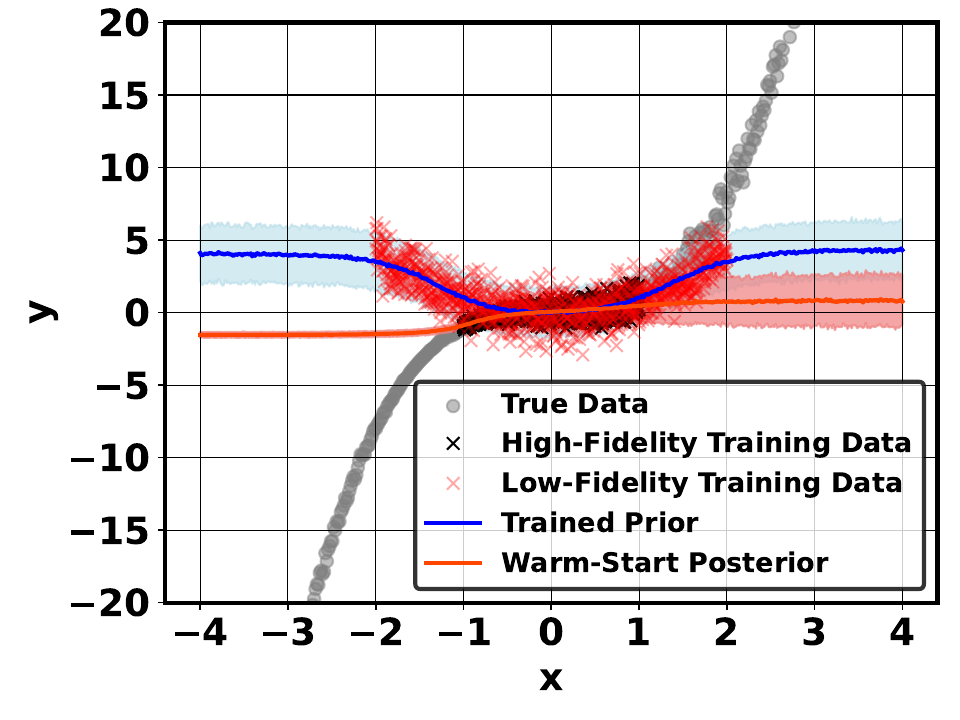}
        \caption{}
        \label{fig:warmstart}
    \end{subfigure}
    \hfill
    \begin{subfigure}[b]{0.49\textwidth}  
        \centering 
        \includegraphics[width=\textwidth]{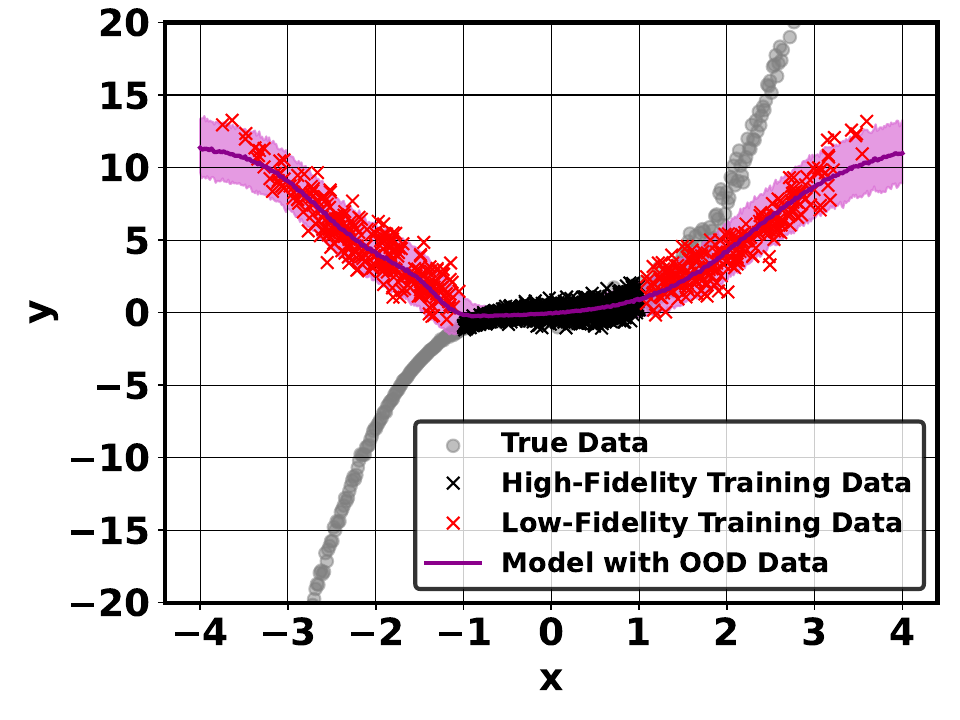}
        \caption{}
        \label{fig:combined}
    \end{subfigure}
    \caption{Demonstration of ``catastrophic forgetting'' when a \gls{BNN} model is (a) first trained to a prior-enforcing dataset and then the true dataset (the warm-start method) and (b) trained on combined dataset.}
    \label{fig:extrap}
\end{figure*}

A demonstration of these principles is presented in Fig.~\ref{fig:extrap}. Here, the underlying true data is the same as for Fig.~\ref{fig:uncertainty-cartoon}. On the other hand, the functional form of the synthetic data is chosen to be 
\begin{equation}
    \hat{y} = x^2 + \eta,
\end{equation}
where $\eta\sim\mathcal{N}(0,0.25^2)$. In Fig.~\ref{fig:warmstart}, it is shown that training on the true data pairs $(\mathbf{x}_i,y_i)_i$ after an initial training phase to the ``prior'' data $(\mathbf{x}_i,\hat{y}_i)_i$ demonstrates an inability to reproduce the prior. However, when the two datasets are concatenated as in Fig.~\ref{fig:combined}, the model is able to more faithfully reproduce the prior. Clearly, there will be a trade-off between the amount of synthetic, low-fidelity, or ``prior'' data included as well as its distance from the true data pairs necessary to reasonably enforce the prior without spoiling the representation learned from the actual data. These trade-offs are explored in Section~\ref{sec:results-extrap}.

To generate \gls{OOD} synthetic data, two methods are explored: a \gls{SBO} approach \cite{moller2021out} and a \gls{NF} approach \cite{durkan2019neural}. In the \gls{SBO} approach, the \gls{OOD} data is generated based on a distance metric computed with respect to all the available data. The method is particularly useful for small datasets and allows controlling the distance between the \gls{OOD} region spanned and the training data. However, the distance computation can become computationally prohibitive when used with large datasets. In the \gls{NF} approach, synthetic data is uniformly generated over a larger domain than the original training data. Here the data spans a hypercube whose edge size is 40\% larger than the spanned amplitude of each feature. The synthetic data that overlaps with training data is subsequently discarded. To evaluate whether a synthetic data point overlaps with the original training data, the likelihood of the training dataset is estimated via a \gls{NF}. Any synthetic data whose likelihood is higher than a given threshold is discarded. The method is summarized in Algorithm~\ref{algo:nfood}. 

\begin{algorithm}
\caption{Out-of-distribution data generation with normalizing flows. $p_{\rm NF}$ denotes the probability predicted by the normalizing flow and $x_i$ denotes the i$^{th}$ element of the dataset $\boldsymbol{D}$.}
\label{algo:nfood}
\begin{algorithmic}[1]
\State Generate $N$ uniformly distributed data points over a space that encompasses the original dataset $\boldsymbol{D}$
\State Train a normalizing flow over the original data $\boldsymbol{D}$ \cite{durkan2019neural}
\State Compute the probability threshold $p_0 = \operatorname{min}_{x_i \in \boldsymbol{D}} p_{\rm NF}(x_i)$
\For{$j=1,N$}
\If{$p_{\rm NF} < p_0$} 
\State reject sample $j$
\Else
\State accept sample $j$
\EndIf
\EndFor
\end{algorithmic}
\end{algorithm}

Unlike \gls{SBO}, the \gls{NF} approach does not allow to control the distance between the synthetic \gls{OOD} dataset and the original dataset, however, it allows fast processing of very large datasets \cite{hassanaly2023uniform}. Here, the likelihood threshold $p_0$ is the minimal likelihood predicted over the true training set, which ensures that the \gls{OOD} region spanned by the synthetic data is strictly disjoint from the original dataset $\boldsymbol{D}$. 

To account for extrapolation uncertainty, the \gls{OOD} dataset generated can be labeled with a label distribution that describes how much uncertainty can be expected if the model is queried outside the original dataset $\boldsymbol{D}$. To simplify the discussion on extrapolation uncertainty, the \gls{OOD} data is labeled as 

\begin{equation}
    \label{eq:oodlabel}
    \chi_{\rm SFS,OOD} \sim \mathcal{N}(\mu_{\rm OOD}, \sigma_{\rm OOD}), 
\end{equation}

where $\mathcal{N}$ denotes a normal distribution with mean $\mu_{\rm OOD}$ being arbitrarily set to 0. The uncertainty in the OOD is set to $\sigma_{\rm OOD} = \operatorname{Var}_{\boldsymbol{D}}(\chi_{\rm SFS})^{1/2}$ which simplifies the implementation and the evaluation (Sec.~\ref{sec:results-extrap}). More complex strategies for generating \gls{OOD} labels could be formulated, such as labeling the synthetic \gls{OOD} data using a \gls{LRM}. Here, an arbitrary marker is used which can primarily serve to detect \gls{OOD} queries (see Sec.~\ref{sec:OODquery}). In Sec.~\ref{sec:results-extrap}, the performance of both synthetic data generation strategies are evaluated with respect to the number of data generated and distance to the training set.

\subsection{Model Specification}
\label{sec:modeling-specification}
In this study, a feed-forward fully connected architecture is considered, similar to that in \cite{yellapantula2021deep}. To account for variability in performance due to model architecture and form, a hyper-parameter tuning study was performed. Specifically, a grid search over all 336 combinations of the following parameters: hidden dimensions $n_h\in\{5,10,15,20\}$, number of hidden layers $n_l\in\{$2, 3, 4$\}$, batch size $M\in\{$256,512,1024,2048,4096,8192,16384$\}$, and learning rate $\eta\in\{$$1\mathrm{e}{-03}$, $1\mathrm{e}{-04}$, $1\mathrm{e}{-05}$, $1\mathrm{e}{-06}$ $\}$ with each model trained for 1500 epochs. Hyperparameter realizations were generated with \texttt{scikit-learn} \cite{kramer2016scikit} and were distributed using the Texas Advanced Computing Center's Launcher utility \cite{Wilson:2014:LSF:2616498.2616534}. The most successful model architecture is summarized in Table~\ref{tab:architecture}. Training took approximately two hours per model on the Eagle computing system at the National Renewable Energy Laboratory. Inference over the roughly two million data points in the validation dataset took on the order of one second. The model was implemented with TensorFlow Probability \cite{dillon2017tensorflow,tensorflow2015-whitepaper} and the code is available in a companion repository \href{https://github.com/NREL/mluq-prop}{https://github.com/NREL/mluq-prop}.

\begin{table}[!hbt]
    \centering
    \begin{tabular}{c|l|l}
         Symbol & Name & Value \\
         \hline
         $n_h$ & Number of units in hidden layer & 20  \\
         $n_l$ & Number of hidden layers & 4 \\
         $M$ & Batch size & 2048 \\
         $\eta$ & Learning rate & $1\mathrm{e}{-04}$ \\
         $\sigma$ & Nonlinear activation function & Sigmoid         
    \end{tabular}
    \caption{Model architecture.}
    \label{tab:architecture}
\end{table}

%% file: results.tex
\section{Results}
\label{sec:results}
In this section, the performance of the \gls{BNN} model is demonstrated through \textit{a priori} testing. As noted in Sec.~\ref{sec:data}, we use the dataset from Ref.~\cite{yellapantula2021deep} where input parameters are generated from the filtered \gls{DNS} data. Model predictions using these input features are compared against filtered \gls{DNS} quantities. We assess the \gls*{BNN} model quality on the validation dataset in Section~\ref{sec:results-predictive}. The predicted aleatoric and epistemic uncertainties and their utility for dataset augmentation are discussed in Section~\ref{sec:results-dist-unc}. Flame contours are modeled and studied in Section~\ref{sec:results-flame}. Finally, the extrapolatory behavior of the model is evaluated in Section~\ref{sec:results-extrap}.

\subsection{Model Predictive Quality}
\label{sec:results-predictive}
\begin{figure}[!htb]
    \centering
    \includegraphics[width=0.65\textwidth]{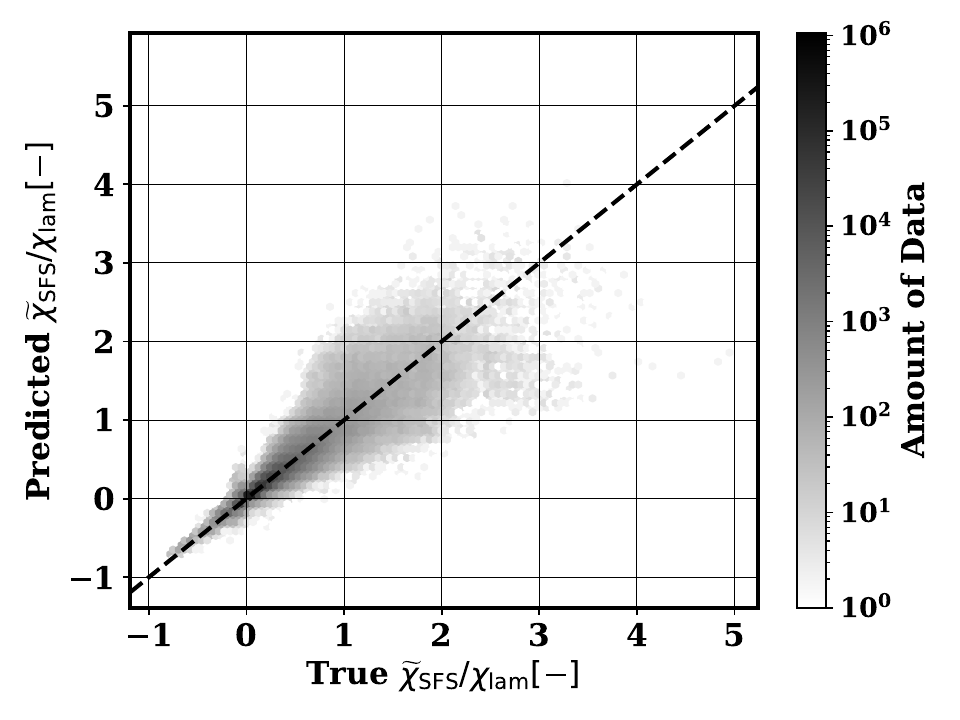}
    \caption{Histogram plot in log-scale showing the mean prediction, $\mathbb{E}_{q(\mathbf{w}|\theta)}\left[ \chi_{\rm SFS} \right]$, from \gls{BNN} model and $\chi_{\rm SFS}$ from \gls{DNS} data. The dashed black line represents a perfect model.}
    \label{fig:hexscatter-mean}
\end{figure}

\begin{figure*}[!htb]
    \centering
    \begin{subfigure}[b]{0.49\textwidth}
        \centering 
        \includegraphics[width=\textwidth]{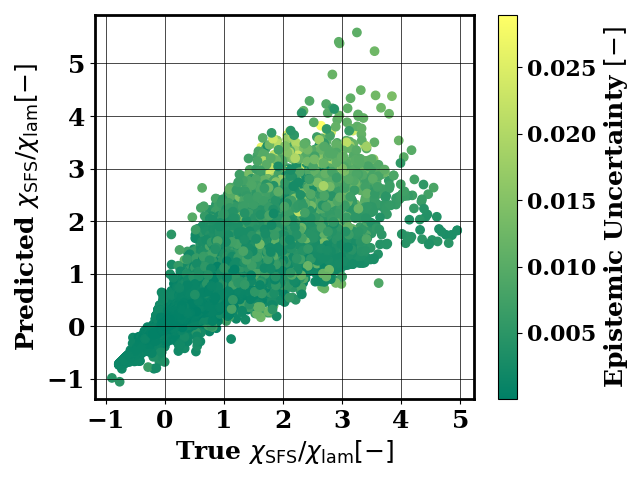}
        \caption{}
        \label{fig:hexscatter-epistemic}
    \end{subfigure}
    \hfill
    \begin{subfigure}[b]{0.49\textwidth}
        \centering 
        \includegraphics[width=\textwidth]{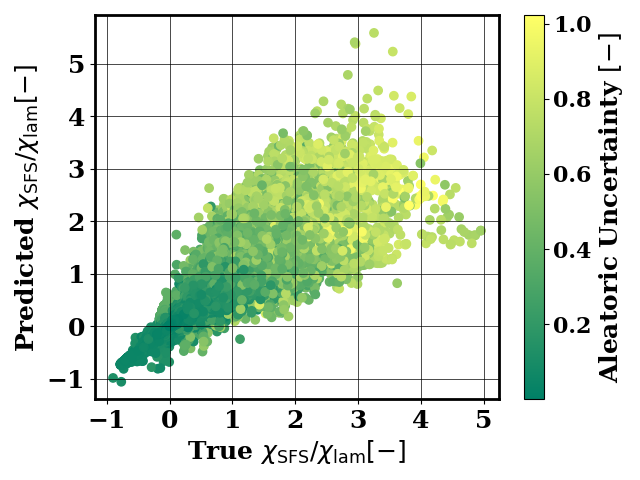}
        \caption{}
        \label{fig:hexscatter-aleatoric}
    \end{subfigure}
    \caption{(a) Scatter plot showing mean predictions for individual data points shaded by the \gls{BNN} model prediction of epistemic uncertainty of $\chi_{\rm SFS}/\chi_{\rm lam}$ and (b) shaded by the \gls{BNN} model prediction of aleatoric uncertainty of $\chi_{\rm SFS}/\chi_{\rm lam}$.}
    \label{fig:hexscatter}
\end{figure*}

The predictive distribution of a \gls{BNN} model trained only on the dataset presented in Sec.~\ref{sec:data}, with no additional synthetic data is shown in Fig.~\ref{fig:hexscatter-mean} and Fig.~\ref{fig:hexscatter}. The mean of the model predictions on the unseen validation data are compared to the true progress variable progress variable dissipation rate as computed by the \gls{DNS} in Fig.~\ref{fig:hexscatter-mean}. An excellent agreement can be observed between the model and data, especially in regions with copious amounts of data. The vast majority of the data clustered along the one-to-one line between the mean prediction values and the true values. At high dissipation rate, the model performance is degraded primarily because the choice of input features does not uniquely map to $\chi_{\rm SFS}$. This is confirmed by Fig.~\ref{fig:hexscatter-aleatoric}
which shows elevated aleatoric uncertainty especially when $\chi_{\rm SFS}/\chi_{\rm lam}$ (where $\chi_{\rm lam}$ is the maximum of the progress variable dissipation rate in the freely propagating laminar flame) exceeds 2. The trained \gls{BNN} model is compared with a \gls{DNN} model trained with the same architecture as the BNN, as well as the physics-based \gls{LRM} \cite{fiorina2005premixed}. The mean-squared error on the test dataset for these models is reported in Table~\ref{tab:mse}. In the case of the \gls{BNN}, the error is computed using the average predictions of the model. We observe that both machine learning approaches outperform the \gls{LRM} (98.5\% error reduction for the \gls{BNN} and 99.4\% error reduction for the \gls{DNN}). The \gls{DNN} slightly outperforms the \gls{BNN} which is unsurprising given that it is trained to only minimize the mean squared error, while the \gls{BNN} also accounts for the \gls{KL} divergence term (Eq.~\ref{eq:elbo}). Overall, the \gls{BNN} allows for obtaining uncertainty estimates while minimally compromising on accuracy.

\begin{table}[]
    \centering
    \begin{tabular}{l|c}
         Model & Mean Squared Error \\
         \hline
         Linear Relaxation Model \cite{fiorina2005premixed} &  $3.98 \times 10^{-1}$ \\
         Deterministic Neural Network \cite{yellapantula2021deep}  & $2.52 \times 10^{-3}$ \\
         Bayesian Neural Network & $5.69 \times 10^{-3}$ \\
    \end{tabular}
    \caption{Comparison of the mean-squared error between various models on the test dataset.}
    \label{tab:mse}
\end{table}

For further evaluation of the model performance, the conditional structure of the progress variable dissipation rate with respect to the input features is inspected. In particular, the performance with respect to three important physical parameters is investigated: the filtered progress variable $\widetilde{C}$, the sub-filter variance of the progress variable $\widetilde{C^{\prime\prime 2}}$, and the resolved progress variable diffusivity $\widetilde{D}_C$. Conditional means were computed with $250$ bins discretizing the input feature domain. Credible intervals are constructed by computing the \gls{BNN} model output across the dataset for 250 realizations of $\mathbf{w}\sim q(\mathbf{w}|\theta)$ and retaining the $(1-\alpha) \times 100\%$ interquantile range of predicted trajectories. In this case, $\alpha=0.1$ corresponding to the $90^{\text{th}}$ percentile. For further discussion of the nuances of predictive, credible, and confidence intervals the reader is referred to \cite{smith2013uncertainty}.

\begin{figure*}[!htb]
    \centering
    \begin{subfigure}[b]{0.45\textwidth}
        \centering
        \includegraphics[width=\textwidth]{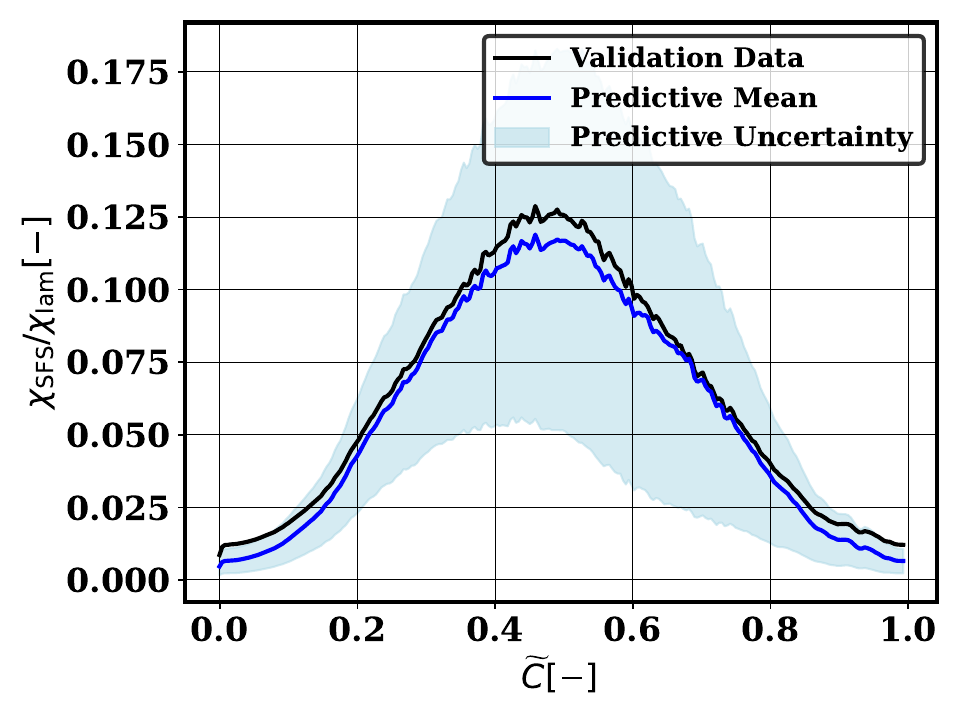}
        \caption{}
        \label{fig:credible-fc}
    \end{subfigure}
    \hfill
    \begin{subfigure}[b]{0.47\textwidth}  
        \centering 
        \includegraphics[width=\textwidth]{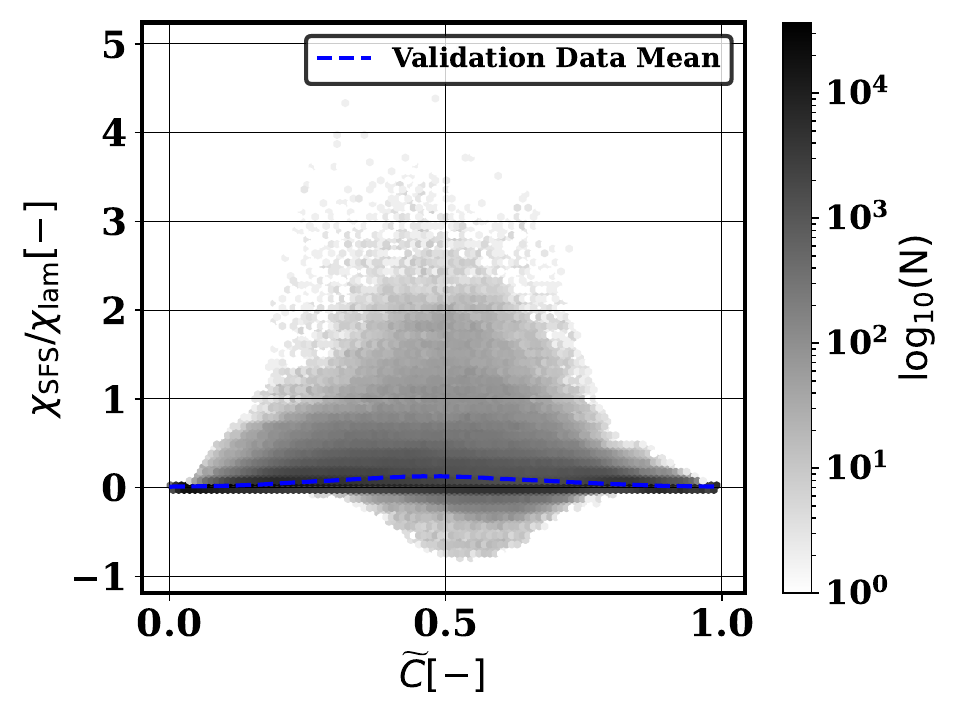}
        \caption{}
        \label{fig:conditional-hex-fc}
    \end{subfigure}
    \vskip\baselineskip
    \begin{subfigure}[b]{0.45\textwidth}
        \centering
        \includegraphics[width=\textwidth]{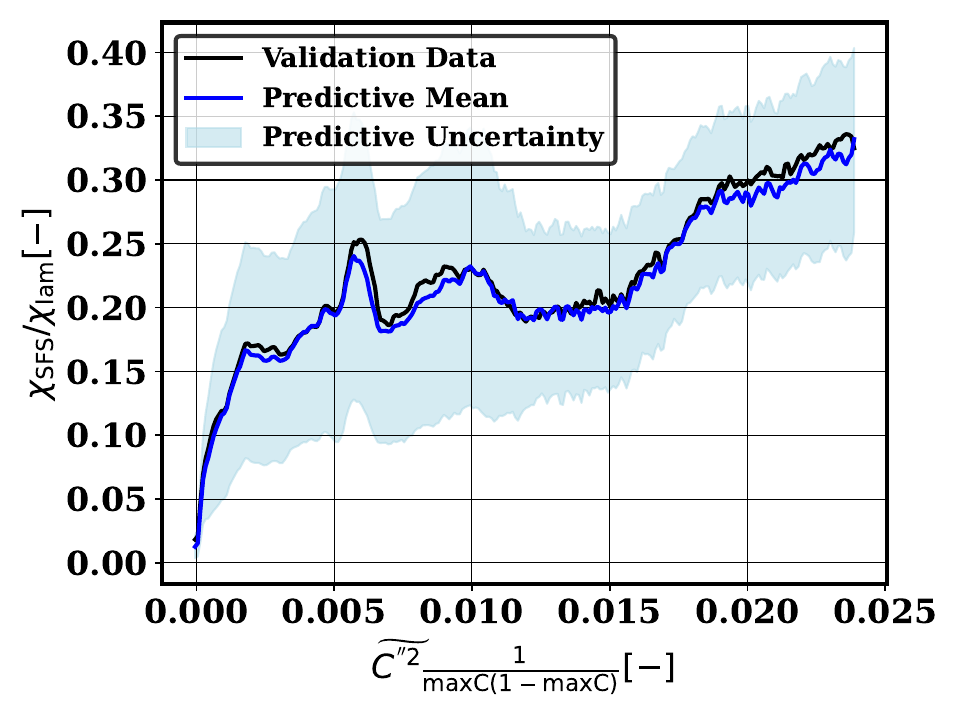}
        \caption{}
        \label{fig:credible-fcvar}
    \end{subfigure}
    \hfill
    \begin{subfigure}[b]{0.47\textwidth}  
        \centering 
        \includegraphics[width=\textwidth]{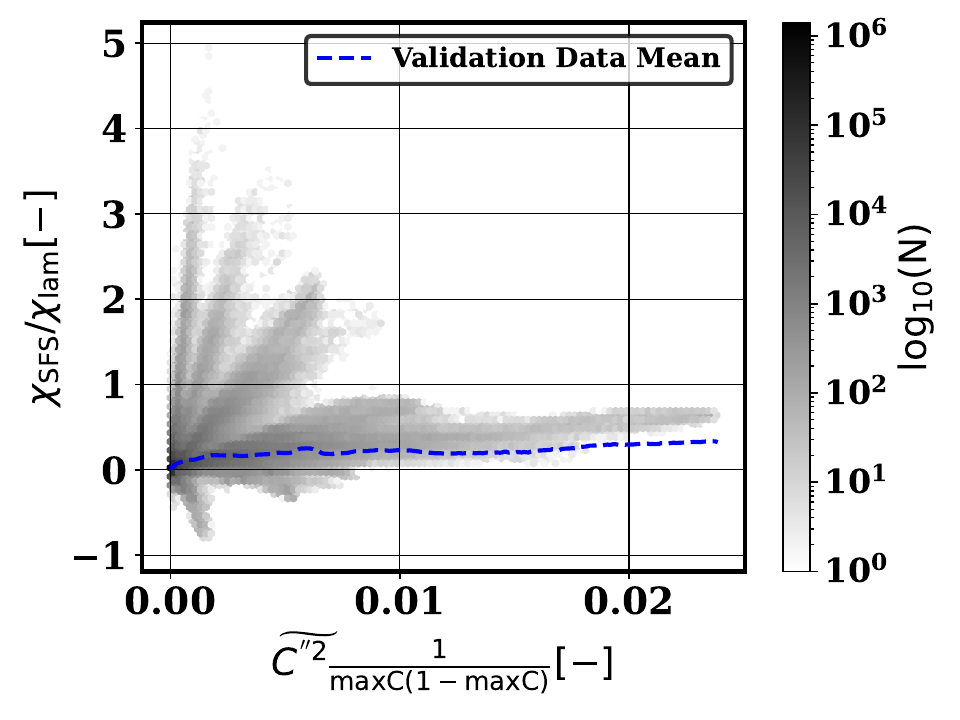}
        \caption{}
        \label{fig:conditional-hex-fcvar}
    \end{subfigure}
    \begin{subfigure}[b]{0.45\textwidth}
        \centering
        \includegraphics[width=\textwidth]{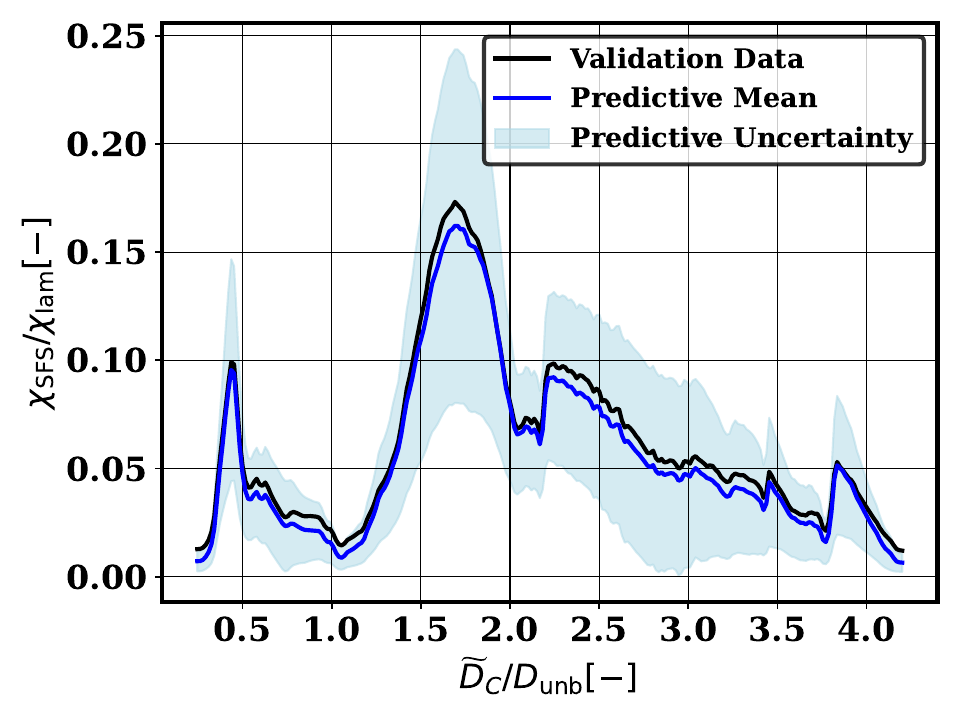}
        \caption{}
        \label{fig:credible-fd}
    \end{subfigure}
    \hfill
    \begin{subfigure}[b]{0.47\textwidth}  
        \centering 
        \includegraphics[width=\textwidth]{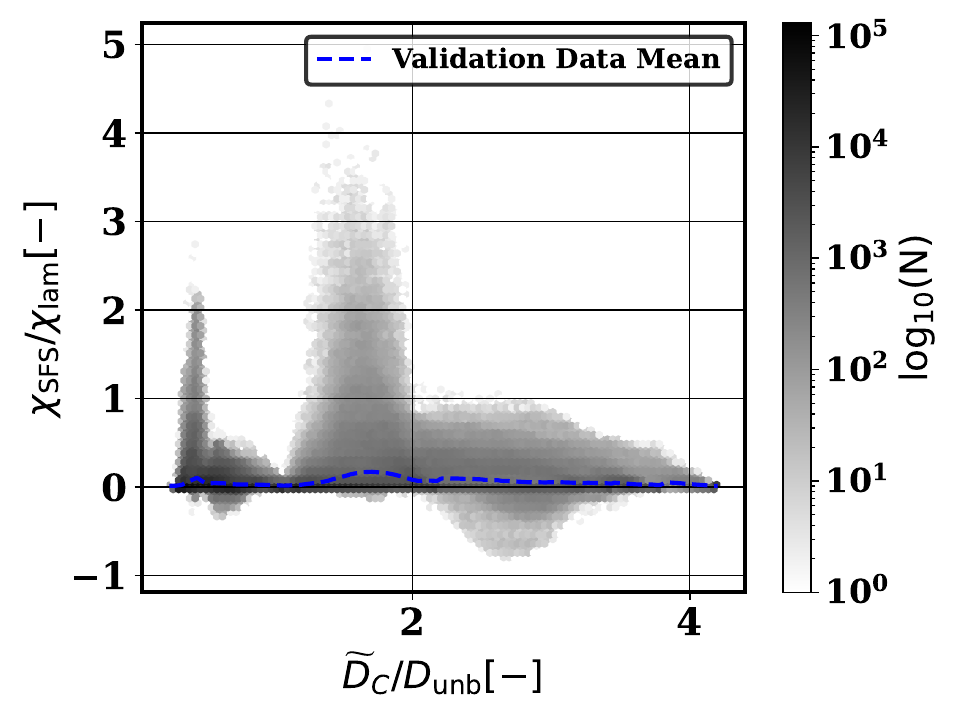}
        \caption{}
        \label{fig:conditional-hex-fd}
    \end{subfigure}
    \caption{Comparison of the model predictive mean and predictive envelope to the validation data mean ((a), (c), (e)) and the distribution of the data conditioned on one dimension of phase space ((b), (d), (f)) for: the filtered progress variable ((a), (b)), the subfilter progress variable variance ((c), (d)), and the filtered progress variable diffusivity ((e), (f)).}
    \label{fig:prediction-hex}
\end{figure*}

The conditional mean profiles displayed in Fig.~\ref{fig:prediction-hex} demonstrate excellent agreement between the data and the model predictive mean. Moreover, any discrepancies, such as those in Fig.~\ref{fig:credible-fcvar} are clearly contained by the predictive envelope. The total predictive uncertainty is dominated by the aleatoric uncertainty as suggested by Fig.~\ref{fig:hexscatter}.

\subsection{Distribution of uncertainties in feature space}
\label{sec:results-dist-unc}

\begin{figure*}[!htb]
    \centering
    \begin{subfigure}[b]{0.49\textwidth}
        \centering
        \includegraphics[width=\textwidth]{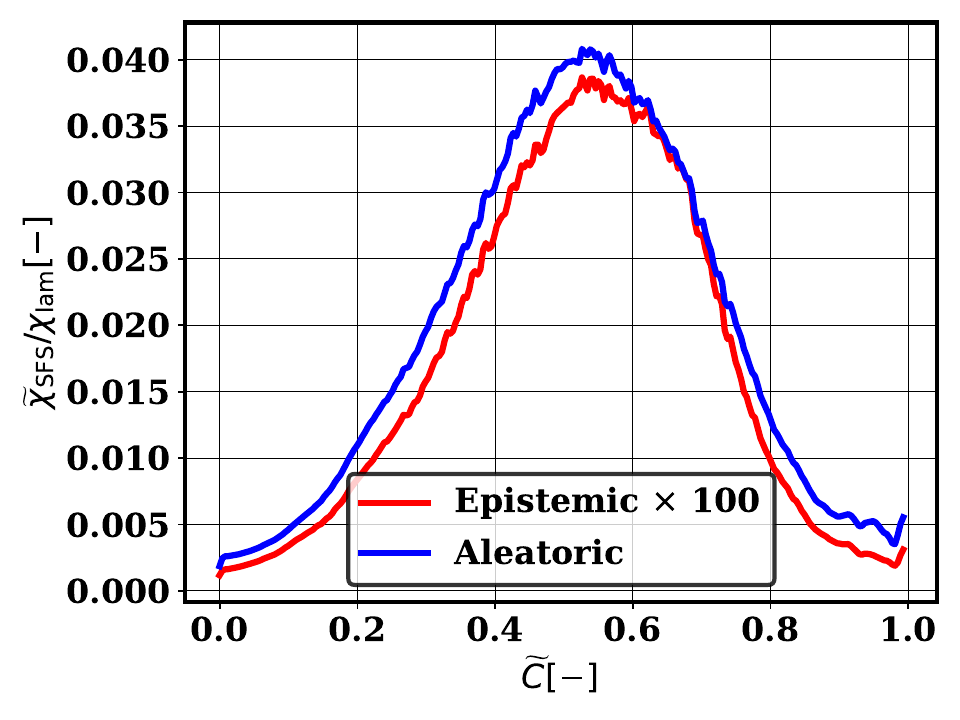}
        \caption{}
        \label{fig:epistemic-fc}
    \end{subfigure}
    \hfill
    \begin{subfigure}[b]{0.49\textwidth}  
        \centering 
        \includegraphics[width=\textwidth]{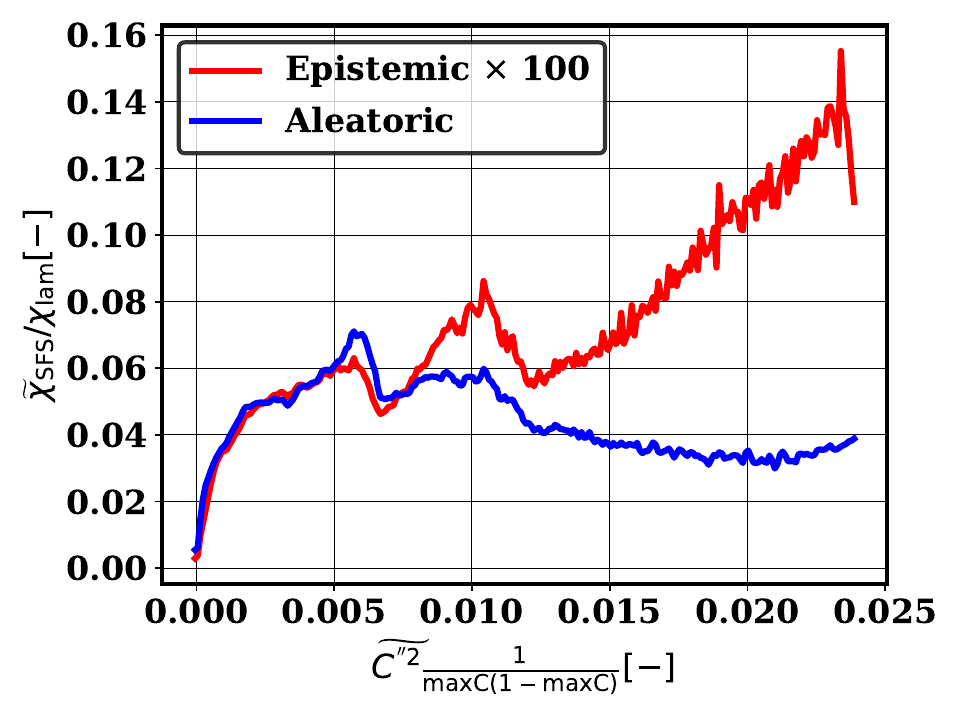}
        \caption{}
        \label{fig:epistemic-fcvar}
    \end{subfigure}
    \vskip\baselineskip
    \begin{subfigure}[b]{0.49\textwidth}
        \centering 
        \includegraphics[width=\textwidth]{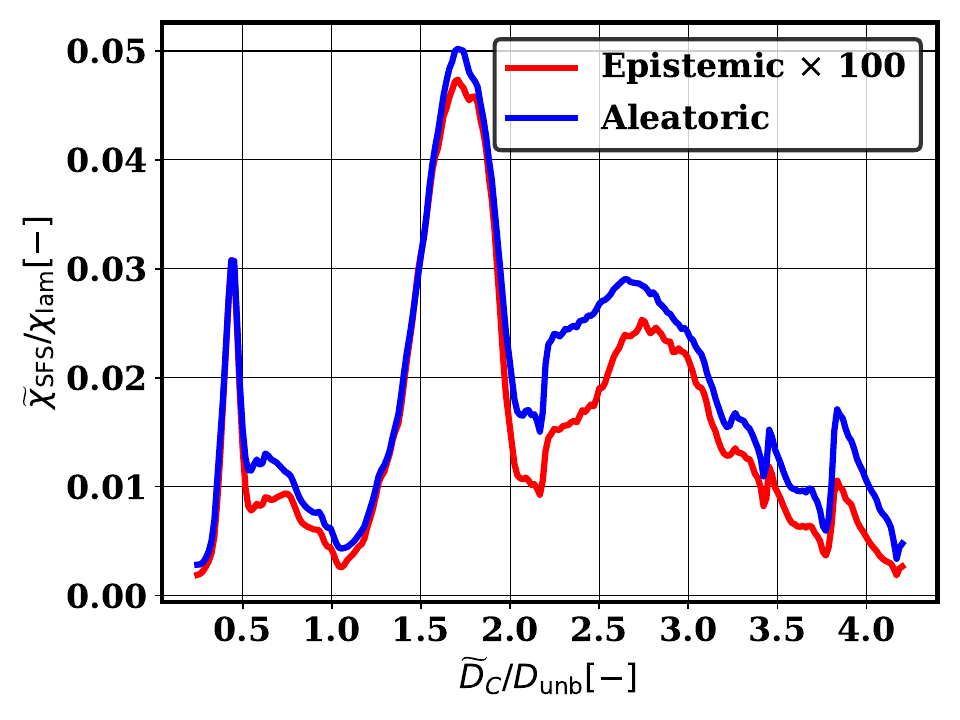}
        \caption{}
        \label{fig:epistemic-fd}
    \end{subfigure}
    \caption{Conditional average of the aleatoric and epistemic uncertainty ($\times$ 100) with respect to (a) filtered progress variable, (b) subfilter progress variable variance, and (c) filtered diffusivity.}
    \label{fig:epistemic-stddev}
\end{figure*}

To estimate and visualize the distribution of epistemic and aleatoric uncertainties, 250 realizations of the model $\mathbf{w}\sim q(\mathbf{w}|\theta)$ are generated and the sample variance estimate of the epistemic uncertainty $\operatorname{Var}_{q(\mathbf{w}|\theta)}(\mathbb{E}[\mathbf{y}\vert \mathbf{x}])$, is computed. The conditional mean of this quantity is computed with respect to the input features and is presented in Fig.~\ref{fig:epistemic-stddev}. The predicted aleatoric uncertainty is computed similarly as $\mathbb{E}_{q_{c}(\mathbf{w}|\theta)}\left(\operatorname{Var}\left[\mathbf{y}\vert \mathbf{x}\right]\right)$ and is also shown in Fig.~\ref{fig:epistemic-stddev}.

First, the epistemic uncertainty is about two orders of magnitude lower than the aleatoric uncertainty, which echos the finding that aleatoric uncertainty dominates the predictive uncertainty (Fig.~\ref{fig:hexscatter}).
The epistemic uncertainty follows the same trends as the ones observed for the aleatoric uncertainty for the conditioning done with respect to the progress variable (Fig.~\ref{fig:epistemic-fc}) and the filtered diffusivity (Fig.~\ref{fig:epistemic-fd}). However, as the progress variable variance increases (Fig.~\ref{fig:epistemic-fcvar}), the distribution of epistemic and aleatoric uncertainty deviate, which is further discussed in Sec.~\ref{sec:discussion}.

\subsection{Spatial distribution of predictions and uncertainties}
\label{sec:results-flame}
\begin{figure*}[!htb]
    \centering
    \begin{subfigure}[b]{0.49\textwidth}
        \centering
        \includegraphics[width=\textwidth]{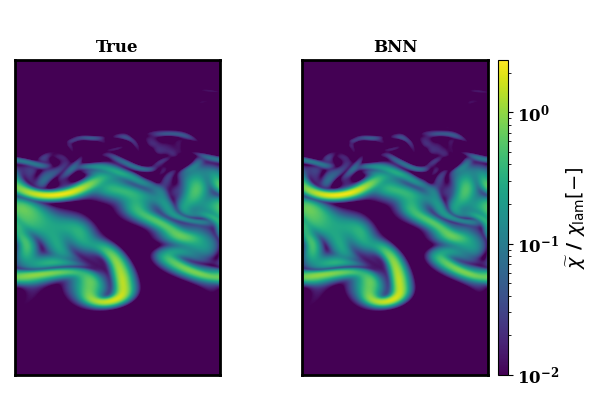}
        \includegraphics[width=\textwidth]
        {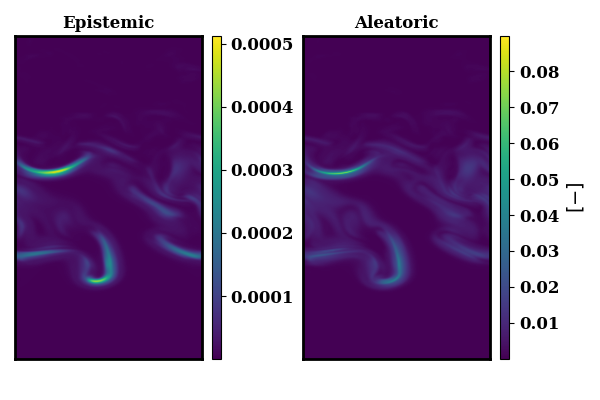}
        \caption{$B_{\rm Le}$ case.}
        \label{fig:Bfw4}
    \end{subfigure}
    \hfill
    \begin{subfigure}[b]{0.49\textwidth}  
        \centering 
        \includegraphics[width=\textwidth]
        {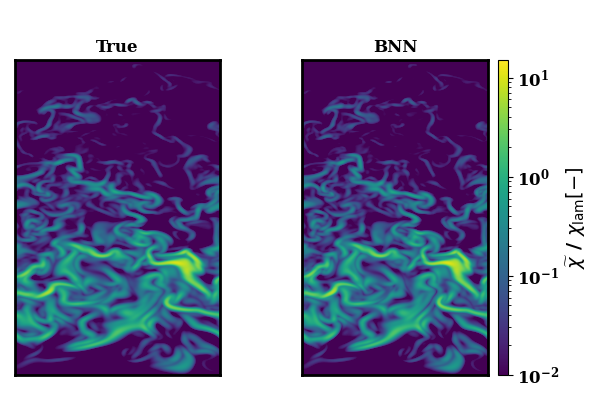}
        \includegraphics[width=\textwidth]{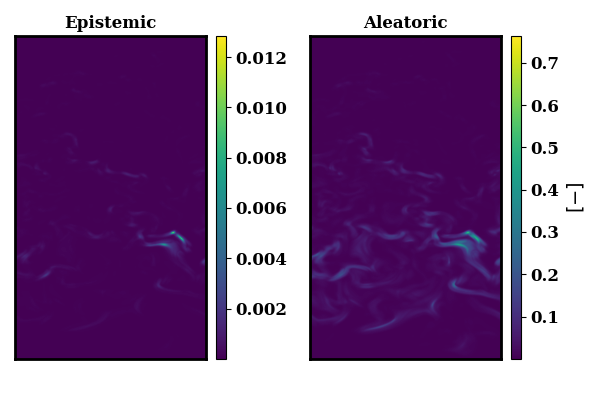}
        \caption{$D_{\rm Le}$ case.}
        \label{fig:Dfw4}
    \end{subfigure}
    \caption{Top: $\widetilde{\chi} / \chi_{\rm lam}$ at a midplane for the ground truth (left) and mean \gls{BNN} prediction (right). Bottom: Epistemic (left) and aleatoric (right) uncertainties associated with the prediction of $\chi_{\rm SFS} / \chi_{\rm lam}$. The bottom of the contours denotes burnt gas and the top of the contours denotes unburnt gas. Results are shown for a filterwidth of 4.}
    \label{fig:FW4}
\end{figure*}

\begin{figure*}[!htb]
    \centering
    \begin{subfigure}[b]{0.49\textwidth}
        \centering
        \includegraphics[width=\textwidth]{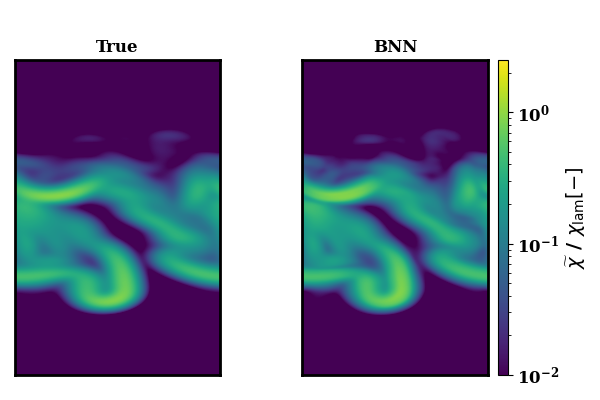}
        \includegraphics[width=\textwidth]
        {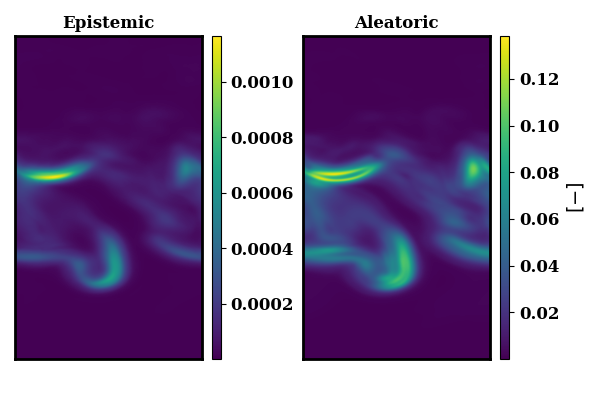}
        \caption{$B_{\rm Le}$ case.}
        \label{fig:Bfw16}
    \end{subfigure}
    \hfill
    \begin{subfigure}[b]{0.49\textwidth}  
        \centering 
        \includegraphics[width=\textwidth]
        {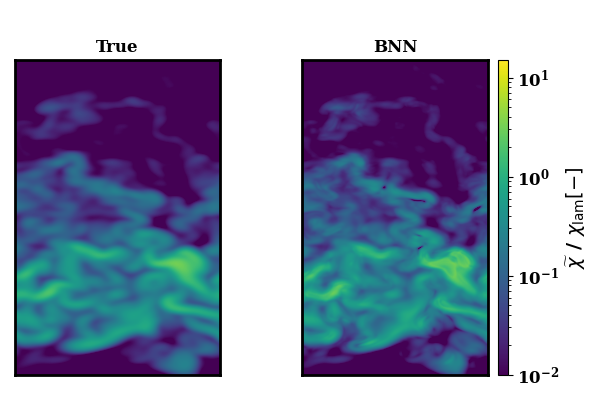}
        \includegraphics[width=\textwidth]{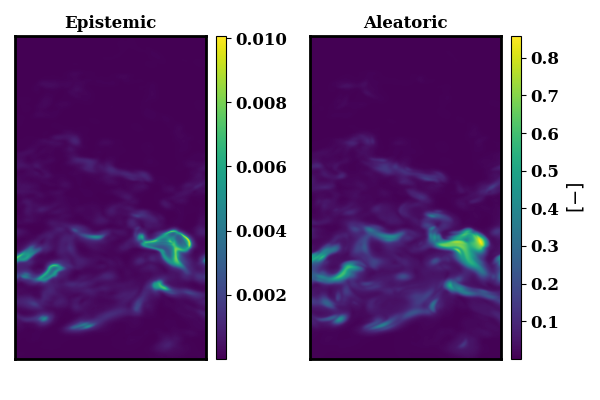}
        \caption{$D_{\rm Le}$ case.}
        \label{fig:Dfw16}
    \end{subfigure}
    \caption{Same as Fig.~\ref{fig:FW4} but with a filterwidth of 16.}
    \label{fig:FW16}
\end{figure*}

The predicted total filtered progress variable dissipation rates, $\widetilde{\chi} = \chi_{\widetilde{C}} + \chi_{\rm SFS}$ and associated uncertainties (epistemic and aleatoric) are visualized for two different flames: an n-heptane flame with Karlovitz number $7$ \cite{savard2017effects} which was not included in the dataset (referred to as $B_{\rm Le}$), and a n-heptane flame with Karlovitz number $256$ which was included in the dataset (referred to as $D_{\rm Le}$). Since the training dataset can only be constructed with a limited set of operating conditions, testing the model outside of those operating conditions evaluates whether the model is only applicable over the conditions included in the training dataset. The prediction results and the ground truth data are shown for filter width 4 (Fig.~\ref{fig:FW4}) and 16 (Fig.~\ref{fig:FW16}).

Figure~\ref{fig:FW4} shows that the \gls{BNN} is in agreement with the filtered \gls{DNS} data, including for the $B_{Le}$ flame which was not included in the dataset. The \gls{BNN} also appropriately captures the effect of the Karlovitz number in that the total filtered progress variable dissipation rate is about one order of magnitude higher for the $D_{\rm Le}$ than for the $B_{\rm Le}$ case. The effect of the filter width is also appropriately captured as shown in Fig.~\ref{fig:FW16}. At higher filter width, the total filtered progress variable dissipation rate peaks at lower values for both flames. 

In line with the results found in Sec.~\ref{sec:results-dist-unc}, the epistemic uncertainty is consistently about 2 orders of magnitude lower than the aleatoric uncertainty. The spatial distribution of epistemic and aleatoric uncertainties are similar as also noted in Sec.~\ref{sec:results-dist-unc}. However, the epistemic uncertainty is more localized, which was also noted earlier (Fig.~\ref{fig:epistemic-fcvar}). 
At larger filter widths, both the epistemic and the aleatoric uncertainty increase, which is the result of a higher amount of information being destroyed by the filtering operation. In the case of the $D_{\rm Le}$ flame with a filter width of 16, $\widetilde{\chi}$ prediction is not as smooth as the ground truth data, especially where the epistemic uncertainty is high. We attribute this to the dataset being insufficiently rich.

\subsection{Extrapolation Behavior}
\label{sec:results-extrap}
In this section, the performance of the model outside of the original dataset distribution $\boldsymbol{D}$ is evaluated. The model's learned representation in the \gls{OOD} regime is mainly driven by the choice of synthetic dataset (Sec.~\ref{sec:modeling-extrapolation}). The role of the synthetic dataset is to enforce an \gls{OOD} behavior without spoiling the performances in distribution. The objective of this section is to understand how should the synthetic dataset be generated. In particular, the choice of \gls{SBO} or \gls{NF} is evaluated, the balance between the size of the synthetic data and the original data is discussed, and the effect of the distance between the synthetic data and the original dataset is described where applicable. A baseline model is first trained with the architecture prescribed in Tab.~\ref{tab:architecture} only using the filtered \gls{DNS} dataset described in Sec.~\ref{sec:data}. The in-distribution performance of the baseline model was previously presented in Sec.~\ref{sec:results-predictive} and Sec.~\ref{sec:results-dist-unc}. The variational posterior of this reference model is denoted by $q_r(\mathbf{w}\vert\theta)$. Comparison models with identical architectures to the reference model are then trained with datasets augmented by various amounts of synthetic data generated as described in Sec.~\ref{sec:modeling-extrapolation}. The variational posterior of these candidate models is denoted $q_c(\mathbf{w}\vert\theta)$.

The normalized $L_2$ error for the first moment outside the data distribution is computed as 

\begin{equation}   
    \label{eq:normErrmom1}
    \|  \mu_{\rm OOD} - \mathbb{E}_{q_{c}(\mathbf{w}|\theta)}\left(\mathbb{E}\left[\mathbf{y}\vert \mathbf{x}\right]\right)\|_2/\sqrt{N},
\end{equation}

where the $L_2$ norm is computed over the set of synthetic data $\boldsymbol{D_{\rm OOD}}$ and $\mathbf{x}$ $N$ is the number of synthetic data points in $\boldsymbol{D_{\rm OOD}}$, and the normalization $1/\sqrt{N}$ accounts for the different synthetic dataset sizes. Likewise, the normalized $L_2$ error for the second moment is computed as

\begin{equation}
    \label{eq:normErrmom2}
    \|\sigma_{\rm OOD}^2 -  \mathbb{E}_{q_{c}(\mathbf{w}|\theta)}\left(\operatorname{Var}\left[\mathbf{y}\vert \mathbf{x}\right]\right)\|_2/\sqrt{N}.
\end{equation}

\begin{figure*}[!htb]
    \centering
    \begin{subfigure}[b]{0.49\textwidth}
        \centering
        \includegraphics[width=\textwidth]{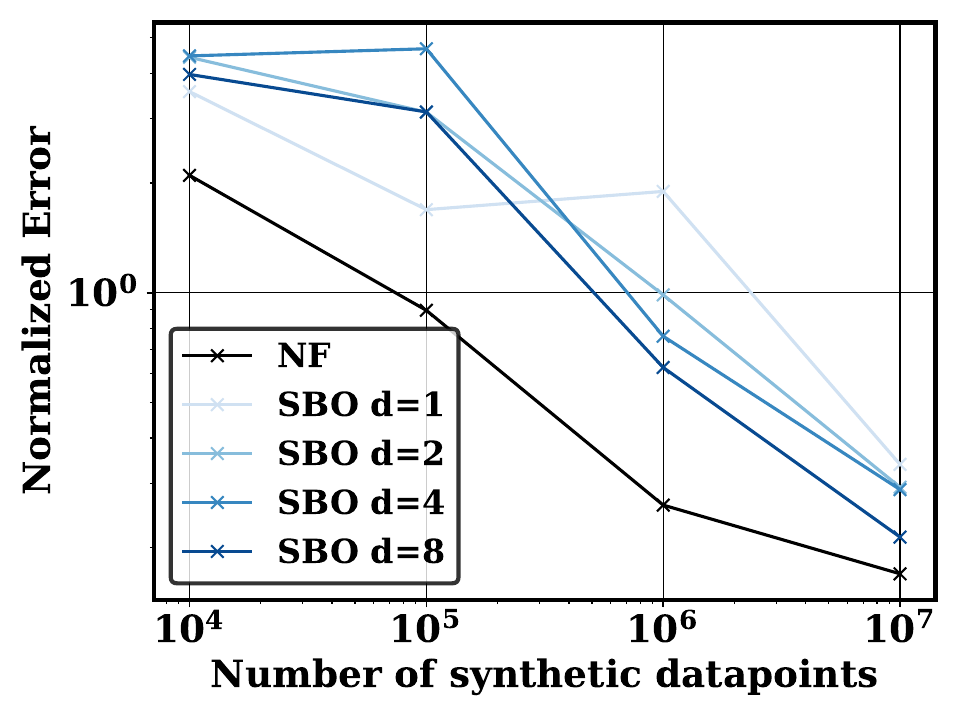}
        \caption{}
        \label{fig:extrap-normalized-mean}
    \end{subfigure}
    \hfill
    \begin{subfigure}[b]{0.49\textwidth}  
        \centering 
        \includegraphics[width=\textwidth]{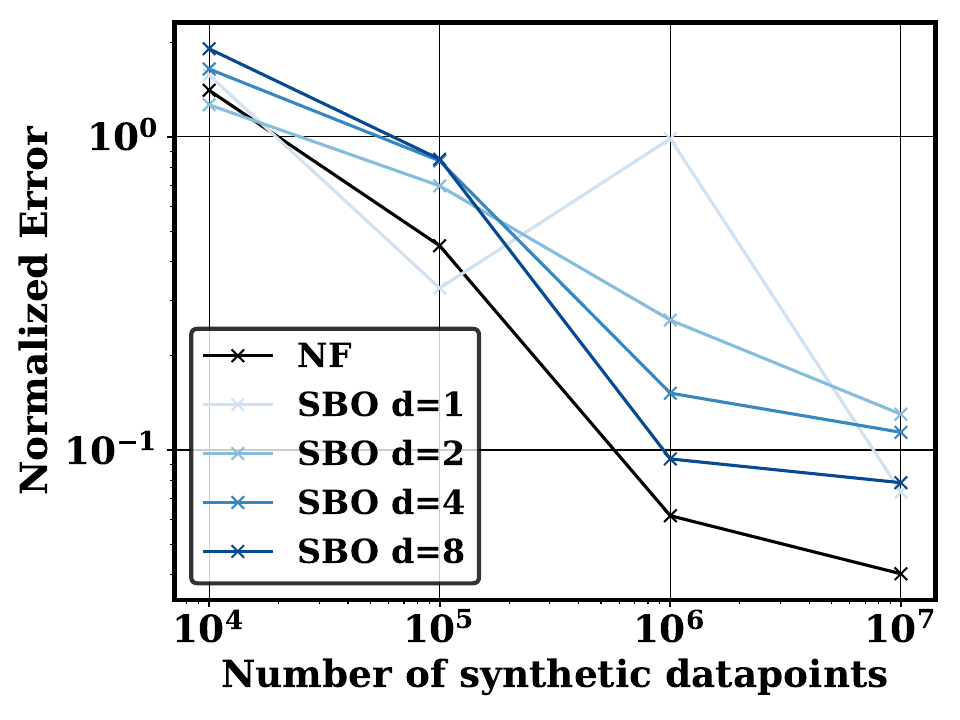}
        \caption{}
        \label{fig:extrap-normalized-std}
    \end{subfigure}
    \caption{Convergence of the normalized $L_2$ error between (a) the model predictive means and the synthetic data in the extrapolation regime and (b) the aleatoric uncertainty estimate and the true aleatoric uncertainty associated with the synthetic data generation.}
    \label{fig:extrap-perf}
\end{figure*}

Convergence plots for these metrics with respect to the amount of synthetic data are presented in Fig.~\ref{fig:extrap-perf}. Unsurprisingly, the higher the amount of synthetic data, the lower the error in the space spanned by the synthetic dataset for both the first and the second moment. It can also be observed that for the same amount of data, generating the \gls{OOD} data with the NF method consistently outperforms all the \gls{SBO} methods. Furthermore, the distance between the synthetic and the in-distribution dataset that can be controlled with the $d$ parameter does not bridge the gap with the NF method. This observation suggests that the difference in performance between NF and \gls{SBO} may be instead due to synthetic dataset distribution.  Unlike \gls{NF}, the \gls{SBO} method does not generate uniformly distributed data in the \gls{OOD} region. In addition, the bounds of the \gls{OOD} region are explicitly specified with the NF method but not in the \gls{SBO}. The gap in performance between the \gls{NF} and the \gls{SBO} methods could be due to undersampled parts of the \gls{OOD} region. Finally, although the error metric converges exponentially fast with respect to the amount of synthetic data, the convergence rates appear to decrease when the amount of synthetic data approaches the amount of in-distribution data.

\begin{figure*}[!htb]
    \centering
    \includegraphics[width=0.49\textwidth]{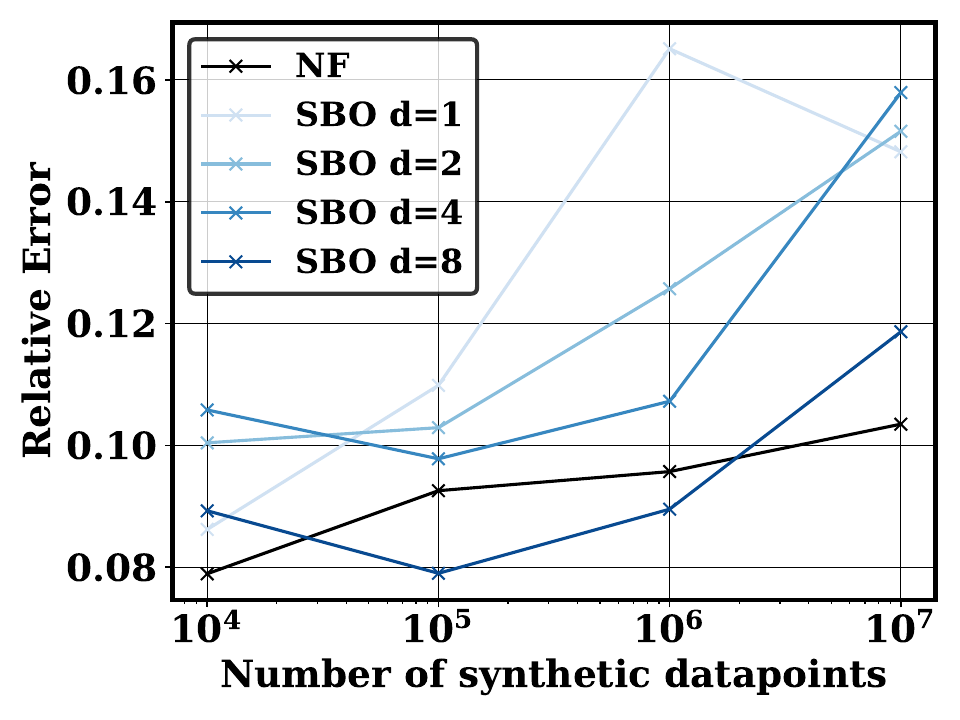}
    \caption{Convergence of the relative $L_2$ error of the predictive mean compared to a reference model on the in-distribution dataset. The models are trained on the original dataset augmented with various amounts of synthetic data.}
    \label{fig:extrap-rel}
\end{figure*}

With the addition of increasing amounts of synthetic data, the model is tasked with representing a larger space and it is natural to expect that the performance should degrade on the original dataset. Figure~\ref{fig:extrap-rel} shows the effect of the amount of synthetic data on the relative $L_2$ error on the original dataset. For a given model $M$, the predictive mean is noted as $P_M = \mathbb{E}_{q_{M, c}(\mathbf{w}|\theta)}\left[\mathbf{y}\vert \mathbf{x}\right]$. The error metric considered is a relative error between the mean predictions of a model trained without synthetic data $P_{\rm ref}$ and a model trained with different amounts of synthetic data $P_{\rm OOD}$, as

\begin{equation}
    \label{eq:normErrmom1-inD}
    \| P_{\rm OOD}  - P_{\rm ref} \|_2 / \| P_{\rm ref} \|_2.
\end{equation}

The relative errors in the predictive mean are shown in Fig.~\ref{fig:extrap-rel}. It is observed that the quality of the predictive mean is affected by the presence of synthetic data. However, the relative error increases at a slower rate than the error reduction in the \gls{OOD} regime. The relative error increases at a faster rate once the number of synthetic data points approaches the size of the in-distribution dataset, especially for the \gls{SBO} data generation. Once more, it appears that the \gls{NF} tends to generate higher quality data.

%% file: discussion.tex
\section{Discussion and Applications}
\label{sec:discussion}

In this section, the uncertainties estimated with the BNN are discussed to describe how they can be used to improve data-driven modeling of closure terms, either to improve the definition of input features, the data collection effort or detect out-of-distribution queries. We discuss the role of the uncertainty decomposition in Section~\ref{sec:discussion-interp}. We discuss how the extrapolation behavior of the BNN can be used for out-of-distribution detection in Section~\ref{sec:OODquery}. Finally, an outlook for \textit{a posteriori} uncertainty propagation through high-fidelity forward simulations is discussed in Section~\ref{sec:discussion-unc-prop}.

\subsection{Interpretation of epistemic and aleatoric uncertainties}
\label{sec:discussion-interp}
For both the BNN (Fig.~\ref{fig:hexscatter}) and the DNN \cite{yellapantula2021deep}, higher errors were primarily observed for large values of $\chi_{\rm SFS}/\chi_{\rm lam}$. There, large aleatoric uncertainties are consistently observed (Fig.~\ref{fig:hexscatter-aleatoric}) which suggests that the input features are insufficiently fine-grained for large filterwidth ratio, or for intense turbulence. In the future, it could be useful to design input features that specifically minimize the aleatoric uncertainty in this region.

The distribution of aleatoric uncertainty with respect to the progress variable (Fig.~\ref{fig:conditional-hex-fc}) suggests that most of the aleatoric uncertainty is located in the burning region of the flame, while reduced uncertainty can be observed either in the fully burnt or unburnt regions. The distribution of aleatoric uncertainty is also skewed towards the high end of the progress variable distribution which may suggest that small-scale variability is mostly controlled by the end of the burning process, as also noted in Ref.~\cite{hassanaly2019ensemble}.

The subfilter variance of the progress variable $\widetilde{C^{\prime\prime 2}}$ may be understood as a marker for the filter width of the \gls{LES}. As the filter width increases, the aleatoric uncertainty rapidly increases which suggests the presence of small-scale variability that cannot be captured with the filtered features (Fig.~\ref{fig:conditional-hex-fcvar}). The aleatoric uncertainty also appears to plateau at high progress variable variance $\widetilde{C^{\prime\prime 2}}$. Since $\widetilde{C^{\prime\prime 2}}$ characterizes small scale variability, it is increasingly difficult for the filtered input features to capture $\chi_{\rm SFS}$. This would result in monotonically increasing aleatoric uncertainty with respect to $\widetilde{C^{\prime\prime 2}}$. However, the plateauing behavior instead suggests that the subgrid scale variability of $\chi_{\rm SFS}$ that can be captured by the filtered input features saturates at medium values of $\widetilde{C^{\prime\prime 2}}$.

Overall, the consistent distribution of epistemic and the aleatoric uncertainties showns in Fig.~\ref{fig:epistemic-stddev} could be a consequence of the near-uniform phase-space sampling done as a pre-processing step (Sec.~\ref{sec:data}). Aleatoric uncertainty is not uniformly distributed over features space, meaning that not all feature regions ``need" the same amount of data. This suggests that the downsampling procedure should attempt to over-represent high-aleatoric uncertainty regimes, instead of uniformly distributing the dataset.

When conditioned on the progress variable sub-filter variance (Fig.~\ref{fig:epistemic-fcvar}), the epistemic and aleatoric uncertainties are initially similarly distributed. However, as the progress variable variance increases, aleatoric uncertainty plateaus, while epistemic uncertainty increases. This observation suggests that highest filter width variances were undersampled in the final dataset. This may be a consequence of the inadequacy of performing uniform-in-phase-space sampling with a clustering approach as discussed in Ref.~\cite{hassanaly2023uniform}.

Finally, it was consistently observed that the aleatoric uncertainty exceeded the epistemic uncertainty (Fig.~\ref{fig:epistemic-stddev}, Fig.~\ref{fig:FW4} and Fig.~\ref{fig:FW16}). Therefore, most of the errors may be attributed to the coarse-graining uncertainty stemming from the filtering operation and the feature selection rather than the lack of data. Albeit small, the epistemic uncertainty could also rapidly increase for a model deployed \textit{a posteriori} in a chaotic system such as a turbulent combustion simulation \cite{hassanaly2019ensemble,hassanaly2019lyapunov}.

\subsection{Using the extrapolation behavior for out-of-distribution query detection}
\label{sec:OODquery}

The extrapolative capabilities enforced with the synthetic dataset can serve two purposes: first, they can ensure that extrapolation is at least as accurate as some low-fidelity model; second, they can serve as a marker for \gls{OOD} queries. In this work, the synthetic data labeling was set arbitrarily. Therefore the trained \glspl{BNN} described in Sec.~\ref{sec:results-extrap} best serve the second purpose: \gls{OOD} query detection. The task of ensuring accuracy OOD is left for future work and would require labeling the synthetic dataset with the low-fidelity model prediction and is left for future work.

\subsubsection{Detection Algorithm}

Using a \gls{BNN} trained with synthetic \gls{OOD} data, the \gls{OOD} query detection method can be done as described in Algo.~\ref{alg:OODdetec}. 

\begin{algorithm}
\caption{Out-of-distribution query detection for input $\mathbf{x}$.}
\label{alg:OODdetec}
\begin{algorithmic}[1]
\State Define a distance threshold $T$
\State Predict $\chi_{\rm SFS, \mu} = \mathbb{E}_{q_{c}(\mathbf{w}|\theta)}\left(\mathbb{E} \left[\mathbf{y}\vert \mathbf{x}\right]\right)$ and $\chi_{\rm SFS, \sigma} = \operatorname{Var}_{q_{c}(\mathbf{w}|\theta)}\left(\mathbb{E}\left[\mathbf{y}\vert \mathbf{x}\right]\right)^{1/2}$ 
\State Compute the distance $d_{\rm OOD} = ||(\chi_{\rm SFS, \mu}, \chi_{\rm SFS, \sigma}) - (\mu_{\rm OOD}, \sigma_{\rm OOD}) ||_2$.
\If{$d_{\rm OOD} > T$}
\State $\mathbf{x}$ is in-distribution
\Else
\State $\mathbf{x}$ is out-of-distribution
\EndIf
\end{algorithmic}
\end{algorithm}

In Algo.~\ref{alg:OODdetec}, the distance $d_{\rm OOD}$ to the arbitrary distribution of labels in the \gls{OOD} region is used as a marker for whether or not an input $\mathbf{x}$ is \gls{OOD}. Clearly the choice of the threshold $T$ impacts efficacy of the method. A systematic method to choose $T$ would depend on the labels adopted for the synthetic \gls{OOD} data and is out-of-scope of this work. In the following, $T$ set to $0.6 \times \sqrt{\mu_{\rm OOD}^2 + \sigma_{\rm OOD}^2}$  which was chosen through a sensitivity analysis (See \ref{sec:appendix-threshold}).

\subsubsection{Results}

\begin{figure*}[!htb]
    \centering
    \begin{subfigure}[b]{0.49\textwidth}
        \centering
        \includegraphics[width=\textwidth]{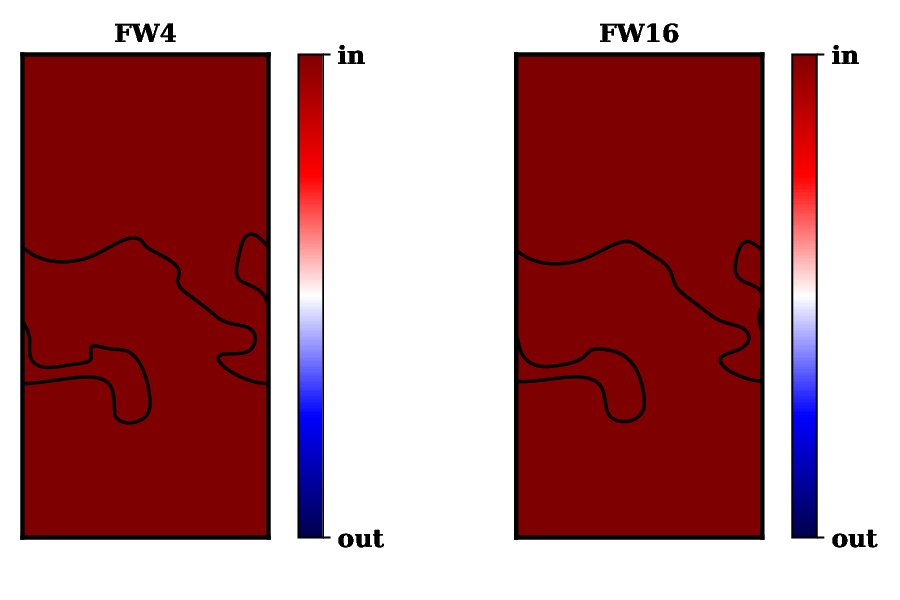}
        \includegraphics[width=\textwidth]{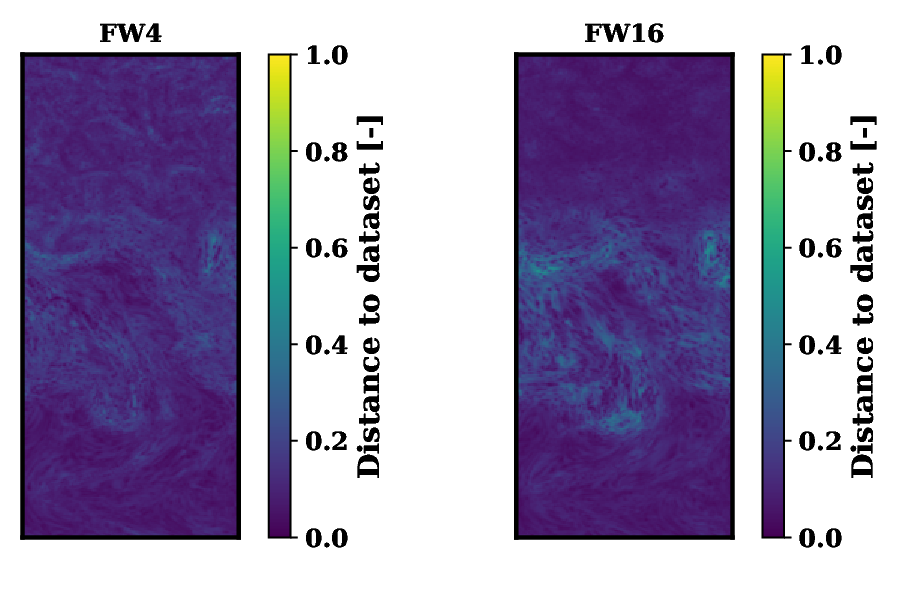}
        \caption{$B_{\rm Le}$ case.}
        \label{fig:BLE_OOD}
    \end{subfigure}
    \hfill
    \begin{subfigure}[b]{0.49\textwidth}
        \centering
        \includegraphics[width=\textwidth]{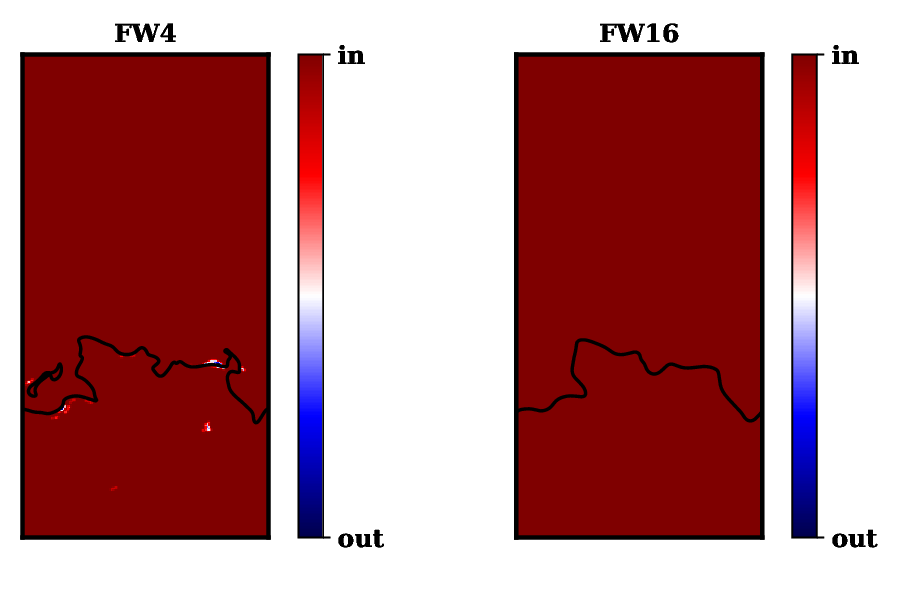}
        \includegraphics[width=\textwidth]{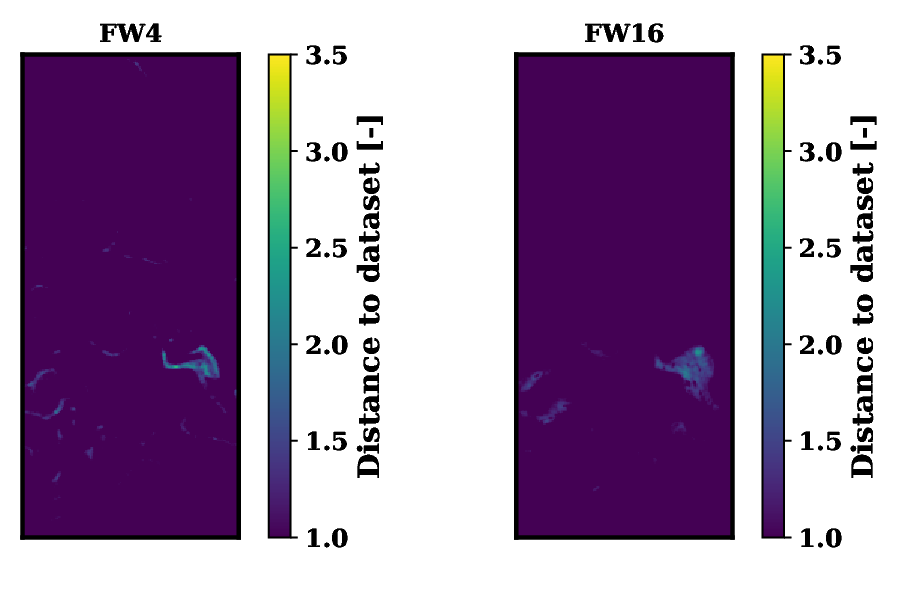}
        \caption{$D_{\rm Le}$ case.}
        \label{fig:DLE_OOD}
    \end{subfigure}
    \caption{Top: in-distribution index predicted by Algo.~\ref{alg:OODdetec} for filterwidth of 4 and 16. The isocontour denotes the flame contour based on the progress variable value. Bottom: distance between the query point and the nearest neighbor of the query point in the training and testing dataset.}
    \label{fig:OOD}
\end{figure*}

The \gls{OOD} query detection is applied to the $D_{Le}$ and $B_{Le}$ flame contours already used in Sec.~\ref{sec:results-flame} for validation of the \gls{BNN} predictions. The \gls{BNN} model used is the one trained with $10^7$ synthetic \gls{OOD} datapoints using the normalizing flow method (Algo. \ref{algo:nfood}). Figure~\ref{fig:OOD} (top) shows the in-distribution index that characterizes whether a \textit{query point} (each point where the \gls{BNN} is queried) is in or out-of-distribution. An in-distribution index is constructed with a value $1$ if $d_{\rm OOD} > T$ and $0$ if  $d_{\rm OOD} < T$. Figure~\ref{fig:OOD} (bottom) shows the distance between the query point and its nearest neighbor in the training and testing datasets. If this distance (hereafter referred to as the \textit{the nearest neighbor metric}) is large, then the query point is far from any data point seen by the BNN during training, and the query point should be considered to be \gls{OOD}. The metric is used to assess whether the \gls{BNN} accurately labels \gls{OOD} query points.

It is observed that the in-distribution index plots and the distance to nearest neighbor metric are in agreement. In particular, the \gls{BNN} successfully labeled query points to be \gls{OOD} when the distance to nearest neighbor is the largest. The \gls{OOD} detection with the \gls{BNN} can be done at the cost of \gls{BNN} inference to compute the empirical averages of $\mathbb{E} \left[\mathbf{y}\vert \mathbf{x}\right]$ (Algo.~\ref{alg:epistemic}) and $\mathbb{E} \left[\mathbf{y}\vert \mathbf{x}\right]$ (Algo.~\ref{alg:aleatoric}) and takes on the order of $\mathcal{O}(10^{-3})$ seconds on a single CPU. The computation of nearest neighbor metric required 108 CPUh (using Intel Xeon Gold Skylake CPUs). Therefore, the \gls{OOD} query detection using a \gls{BNN} is computationally advantageous compared with more naive methods.

Figure~\ref{fig:BLE_OOD} (top) shows that the none of query points from the $B_{Le}$ flame case could be considered out-of-distribution. This is a surprising result given that the $B_{Le}$ flame was not included in the training dataset. Nevertheless, the phase-space spanned by the $B_{Le}$ flame is similar to the training and testing dataset. This result highlights that good predictive performances of data-based closure models on cases not included in the training dataset does not test the extrapolative abilities of data-based closure models. Instead, it may test whether the dataset was sufficiently large so that the data-based model can be used on reacting flows not included in the training dataset.

In turn, Fig.~\ref{fig:DLE_OOD} (top) suggests that the case with filterwidth of 4 for the $D_{Le}$ flame contains \gls{OOD} query points. This is, again, a surprising result given that the $D_{Le}$ flame was included in the training dataset. The nearest neighbor metric (Fig.~\ref{fig:DLE_OOD}, bottom) confirms that this finding is reasonable since the \gls{OOD} query points are indeed far from any training or testing data points. This effect could be explained by the fact that the training data was preprocessed with an approximate uniform-in-phase-space sampling to eliminate redundant datapoints. However the method used in Ref.~\cite{yellapantula2021deep} has been found to be inaccurate since it tends to remove too many rare datapoints \cite{hassanaly2023uniform}. Fortunately, the \gls{OOD} query is more prominent for the filterwidth 4 which is also where the contribution of $\chi_{\rm SFS}$ is negligible compared to $\chi_{\widetilde{C}}$. This explains why the \gls{OOD} queries did not affect the reconstruction of $\widetilde{\chi}$ in this work (Sec.~\ref{sec:results-flame}) or in Ref.~\cite{yellapantula2021deep}.

\subsection{Towards model uncertainty propagation}
\label{sec:discussion-unc-prop}
For the purpose of decision-making, predictive simulations need to be paired with appropriate uncertainty estimates that reflect various uncertainty forms. Several efforts have shown that it was possible to estimate uncertainty due to boundary conditions \cite{masquelet2017uncertainty,khalil2015uncertainty} and the closure model adopted for LES \cite{khalil2015uncertainty,mueller2018model} and RANS \cite{margheri2014epistemic}. Model form uncertainty is typically represented by the variation a few (at most 3 in \cite{khalil2015uncertainty}) model parameters, the ones that appear in physics-based closure models \cite{smagorinsky1963general,spalart1992one,launder1974application} or that are used to superimpose physics-based models \cite{mueller2018model}. Here, a key complication is that the number of uncertain parameters is the number of weights (typically $\mathcal{O}(10^3)$) in the neural network. To enable uncertainty propagation, a key first step would be to reduce the number of uncertain weights. This task is left for future work.

%% file: conclusion.tex
\section{Conclusions}
\label{sec:conclusion}
This work presented the first demonstration of a Bayesian neural network approach for data-driven closure modeling equipped with aleatoric and epistemic uncertainty estimates. The main focus was on modeling subfilter progress variable dissipation rate, but the methods could be applied in general to closure modeling. \textit{a priori} tests showed a good mean prediction of the subfilter progress variable dissipation rate, which suggests that including uncertainty estimates during training does not adversely affect the model accuracy. 

Overall, the aleatoric uncertainty was found to outweigh the epistemic uncertainty. Epistemic and aleatoric uncertainty were found to similarly vary over phase-space, with the exception of large progress variable regions. This is likely a consequence of the approximate uniform-in-phase-space data sampling procedure. These findings motivate the use of non-uniform data selection which could compensate for local pockets of high aleatoric uncertainty in phase space. 

A strategy for enforcing a specific \gls{OOD} behavior was proposed using synthetic data generation. It was found that using a uniform in-phase space data generation led to the best performances and that the performance in-distribution did not degrade until the synthetic dataset size approached that of the original dataset. The \gls{OOD} behavior can be used as a marker that efficiently tests whether the inputs passed to the BNN are far or not from the data-distribution. It was found that the strategy proposed is computationally efficient for identifying \gls{OOD} queries.

Future work will include propagation of epistemic uncertainty through \gls{LES} codes and analysis of the \textit{a posteriori} uncertainty as well as dimension reduction strategies to efficiently propagate the uncertainty without sampling the high dimensional weight-space. Additionally, future work will explore the application of active learning or \gls{OED} to inform future data collection and both \textit{a priori} and \textit{a posteriori} analysis of the model before and after the informed data collection.

%% file: credit.tex
\section{CRediT Authorship Contribution Statement}
GP: Methodology, Investigation, Software, Writing - original draft. MH: Conceptualization, Methodology, Investigation, Software, Supervision, Visualization, Writing - review \& editing. SY: Conceptualization, Data curation, Funding acquisition, Supervision, Writing - review \& editing.

%% file: acknowledgements.tex
\section{Acknowledgements}
\label{sec:acknowledgements}
The authors would like to thank Bruce Perry and Michael Mueller for fruitful discussions.

This material is based upon work supported by the U.S. Department of Energy, Office of Science, Office of Advanced Scientific Computing Research, Department of Energy Computational Science Graduate Fellowship under Award Number DE-SC0021110. This work was authored in part by the National Renewable Energy Laboratory (NREL), operated by Alliance for Sustainable Energy, LLC, for the U.S. Department of Energy (DOE) under Contract No. DE-AC36-08GO28308.
This work was supported as part of DEGREES, an Energy Earthshot Research Center (EERC) funded by the U.S. Department of Energy, and by DOE's Advanced Scientific Computing Research (ASCR) program. SY was supported by the U.S. Department of Energy Office of Energy Efficiency and Renewable Energy Vehicle Technologies Office (VTO).
The research was performed using computational resources sponsored by the Department of Energy's Office of Energy Efficiency and Renewable Energy and located at the National Renewable Energy Laboratory. The views expressed in the article do not necessarily represent the views of the DOE or the U.S. Government. The U.S. Government retains and the publisher, by accepting the article for publication, acknowledges that the U.S. Government retains a nonexclusive, paid-up, irrevocable, worldwide license to publish or reproduce the published form of this work, or allow others to do so, for U.S. Government purposes. This report was prepared as an account of work sponsored by an agency of the United States Government. Neither the United States Government nor any agency thereof, nor any of their employees, makes any warranty, express or implied, or assumes any legal liability or responsibility for the accuracy, completeness, or usefulness of any information, apparatus, product, or process disclosed, or represents that its use would not infringe privately owned rights. Reference herein to any specific commercial product, process, or service by trade name, trademark, manufacturer, or otherwise does not necessarily constitute or imply its endorsement, recommendation, or favoring by the United States Government or any agency thereof. The views and opinions of authors expressed herein do not necessarily state or reflect those of the United States Government or any agency thereof.

%% file: appendix.tex
\appendix
\label{sec:appendix}

\input{appendix-12d}

\input{appendix-uncertainty}

\input{appendix-isoprior}

\input{appendix-threshold}

%% file: appendix-12d.tex
\section{Effect of Additional Input Parameters}
\label{sec:appendix-12D}

\begin{figure}[!htb]
    \centering
    \includegraphics[width=0.45\textwidth]{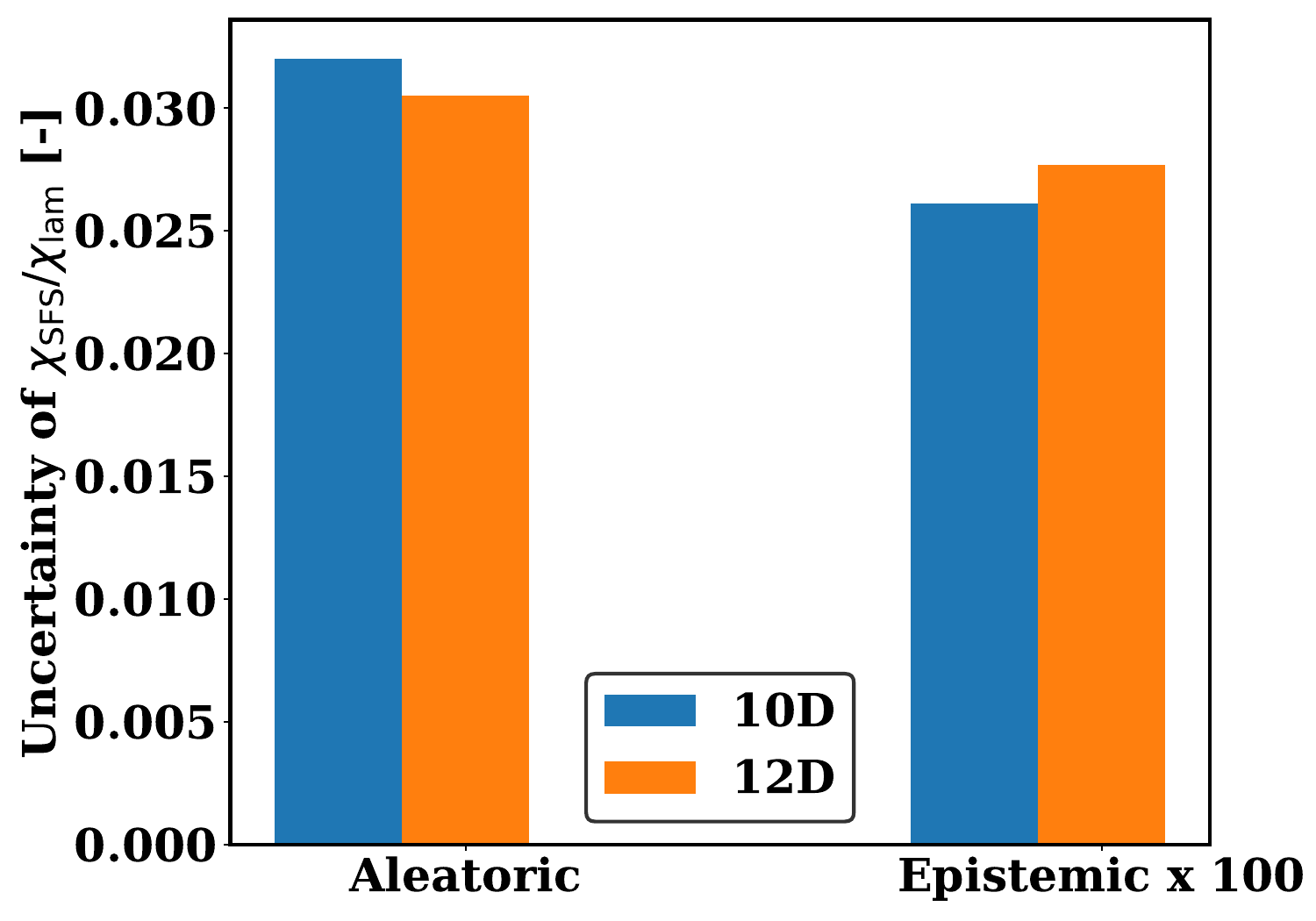}
    \caption{Bar plot of the average epistemic and aleatoric uncertainties over the test dataset for the 10D BNN (blue bar) and the 12D BNN (orange bar).}
    \label{fig:10d-12d}
\end{figure}

To illustrate the effect of choice of input features on the uncertainties estimated, a BNN that uses 12 dimensions (referred to as 12D BNN) was trained with the following two additional features: 1) $\widetilde{\dot{\omega_C}}$, the Favre-filtered reaction source term of the progress variable; and 2) $\nabla \widetilde{T} \cdot \nabla \widetilde{C}$ where $\widetilde{C}$ is the Favre-filtered temperature. Figure~\ref{fig:10d-12d} shows the effect of expanding the feature space on the average aleatoric and epistemic uncertainties over the test dataset, between the 12D BNN and the original BNN trained with 10 input features (referred to as the 10D BNN). None of the datasets are augmented with synthetic OOD data.

When augmenting the input feature space dimension, the aleatoric uncertainty decreases. This can be explained by the fact that the additional input features reduce the loss of information in the filtered dataset. However, using a higher dimensional input space also increases the epistemic uncertainty. This phenomenon can be explained by the fact that data density is higher in the low-dimensional input space, thereby mitigating statistical errors that result from the lack of data.

%% file: appendix-uncertainty.tex
\section{Computation of Uncertainties}
\label{sec:appendix-algo}
In this appendix, algorithms for the computation of a predictive distribution (Alg.~\ref{alg:predictive}),  the epistemic uncertainty (Alg.~\ref{alg:epistemic}) and the aleatoric uncertainty (Alg.~\ref{alg:aleatoric}) are presented. Overall, the \gls{BNN} allows for fast evaluation of the samples of weights, but appropriate averaging is needed to differentiate between epistemic and aleatoric uncertainties.

To compute the predictive distribution, we must both sample realizations of the weights (accounting for epistemic uncertainty) and sample from the resultant random variable (accounting for the aleatoric uncertainty). The can be achieved as a double for loop, with a third, outer loop describing the process of generating predictions across a set of input values. Naively, both inner sampling loops are achieved with Monte Carlo sampling and are subject to the standard $\mathcal{O}(1/\sqrt{n})$ convergence rates, where $n$ is the number of samples.

\begin{algorithm}
\caption{Generate the predictive distribution for a collection of input data $\{\mathbf{x}_i\}^N_i$, given a \gls{BNN} posterior $q(\mathbf{w}\vert\theta)$, and a number of epistemic $N_e$ and aleatoric samples $N_a$.}
\label{alg:predictive}
\begin{algorithmic}[1]
\For{$i=1,\dots, N$}
    \For{$j=1,\dots, N_e$}
        \State Sample $\mathbf{w}_j\sim q(\mathbf{w}\vert\theta)$
        \State Compute \gls{BNN} prediction of $\mathbf{y}_{\boldsymbol{\mu}}$ and $\mathbf{y}_{\boldsymbol{\sigma}}$
        \For{$k=1,\dots, N_a$}
            \State Sample $\mathbf{y}\sim\mathcal{N}(\mathbf{y}_{\boldsymbol{\mu}},\operatorname{diag}(\mathbf{y}_{\boldsymbol{\sigma}}))$ and append to collection of predictions
        \EndFor
    \EndFor
\EndFor
\end{algorithmic}
\end{algorithm}

Computation of the epistemic uncertainty relies on the parameterization chosen for \gls{BNN} output. In the case of a Kendall and Gal~\cite{kendall2017uncertainties} style architecture, the model mean is directly parameterized as one of the model outputs. This fact may be exploited to avoid a sample average computation of the average. A sample variance calculation may then be performed on the collection of mean predictions. The Monte Carlo sampling and sample variance calculation is subject to the error prescribed by the Central Limit Theorem, again $\mathcal{O}(1/\sqrt{N_e})$

\begin{algorithm}
\caption{Compute an estimate of the epistemic uncertainty associated with an input datum $\mathbf{x}$, given a \gls{BNN} posterior $q(\mathbf{w}\vert\theta)$, and a number of epistemic samples $N_e$.}
\label{alg:epistemic}
\begin{algorithmic}[1]
\For{$j=1,\dots, N_e$}
    \State Sample $\mathbf{w}_j\sim q(\mathbf{w}\vert\theta)$
    \State Compute \gls{BNN} prediction of $\mathbf{y}_{\boldsymbol{\mu}}$ corresponding to $\mathbb{E}[\mathbf{y}\vert\mathbf{x}]$ for the weights $\mathbf{w}_j$ and append to collection $\{\mathbf{y}_{\boldsymbol{\mu}_i}\}^{N_e}_i$
\EndFor
\State Compute the sample variance of collection $\{\mathbf{y}_{\boldsymbol{\mu}_i}\}^{N_e}_i$ corresponding to $\operatorname{Var}_{q(\mathbf{w}|\theta)}(\mathbb{E}[\mathbf{y}\vert \mathbf{x}])$
\end{algorithmic}
\end{algorithm}

Computation of the aleatoric uncertainty similarly relies on the parameterization of the model outputs. With the variance, or standard deviation, as one of the model outputs, one may directly obtain this from a model evaluation after sampling the weights $\mathbf{w}\sim q(\mathbf{w}\vert\theta)$. A sample average calculation of this collection then returns an estimate of the aleatoric uncertainty. Once more, the sample mean calculation is subject to the convergence properties of the Central Limit Theorem, $\mathcal{O}(1/\sqrt{N_a})$.

\begin{algorithm}
\caption{Compute an estimate of the aleatoric uncertainty associated with an input datum $\mathbf{x}$, given a \gls{BNN} posterior $q(\mathbf{w}\vert\theta)$, and a number of aleatoric samples $N_a$.}
\label{alg:aleatoric}
\begin{algorithmic}[1]
\For{$j=1,\dots, N_a$}
    \State Sample $\mathbf{w}_j\sim q(\mathbf{w}\vert\theta)$
    \State Compute \gls{BNN} prediction of $\mathbf{y}_{\boldsymbol{\sigma}}$ corresponding to $\operatorname{Var}(\mathbf{y} \vert \mathbf{x})$ for the weights $\mathbf{w}_j$ and append to collection $\{\mathbf{y}_{\boldsymbol{\sigma}_i}\}^{N_a}_i$
\EndFor
\State Compute the mean of collection $\{\mathbf{y}_{\boldsymbol{\sigma}_i}\}^{N_a}_i$ corresponding to $\mathbb{E}_{q(\mathbf{w}|\theta)} \left[\operatorname{Var}(\mathbf{y} \vert \mathbf{x})\right]$
\end{algorithmic}
\end{algorithm}

%% file: appendix-isoprior.tex
\section{Effect of an Isotropic Gaussian Prior on Prediction Quality}
\label{sec:appendix-isoprior}
\begin{figure*}[!htb]
    \centering
    \begin{subfigure}[b]{0.49\textwidth}
        \centering
        \includegraphics[width=\textwidth]{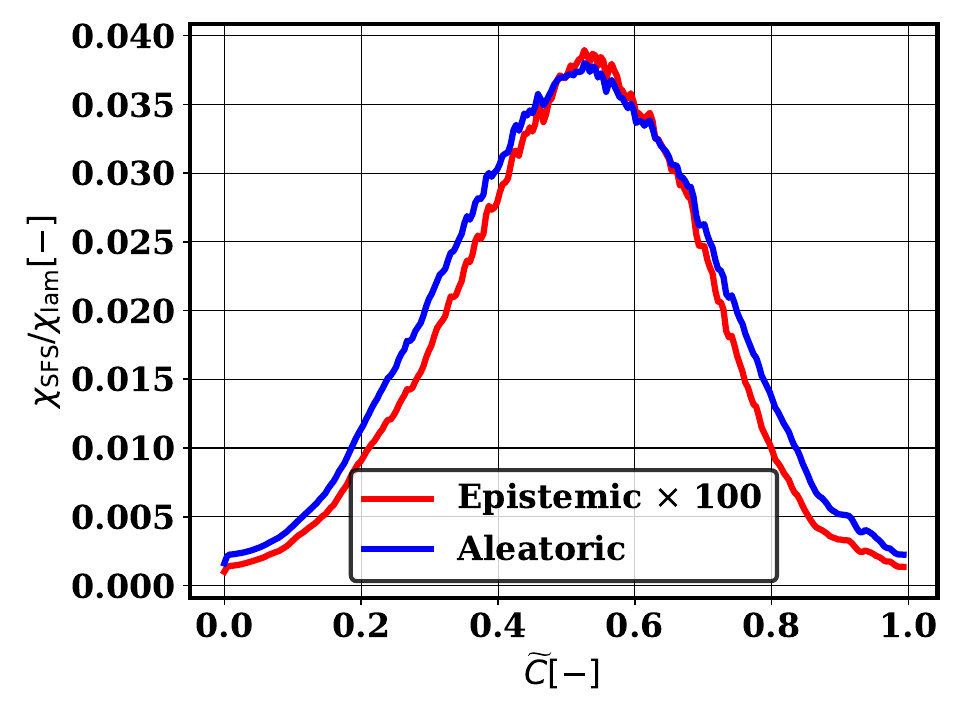}
        \caption{}
        \label{fig:epistemic-fc-isoprior}
    \end{subfigure}
    \hfill
    \begin{subfigure}[b]{0.49\textwidth}  
        \centering 
        \includegraphics[width=\textwidth]{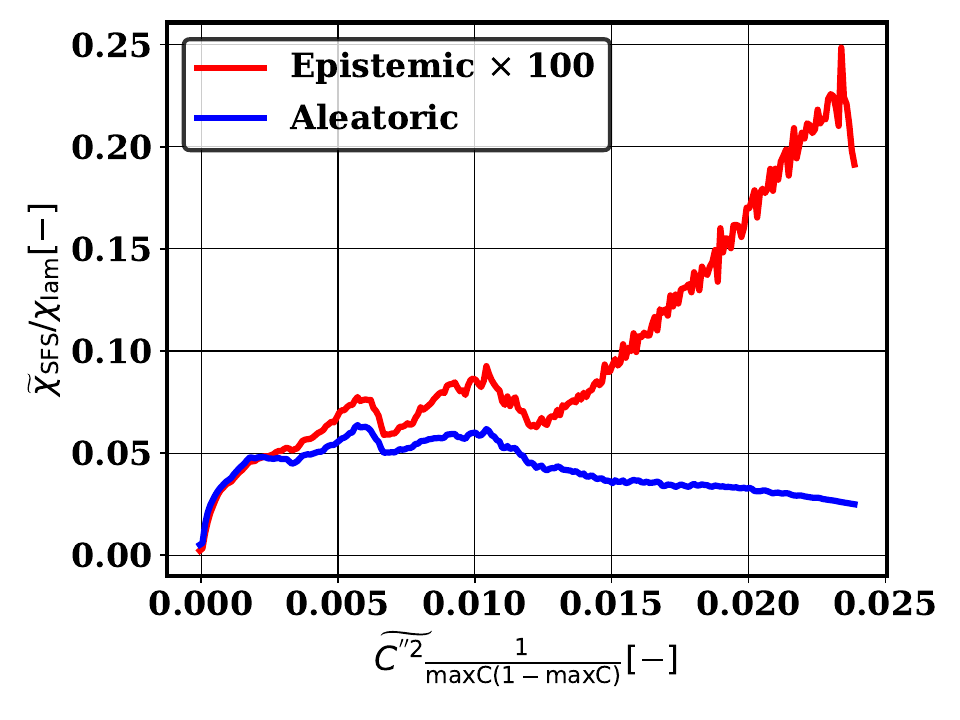}
        \caption{}
        \label{fig:epistemic-fcvar-isoprior}
    \end{subfigure}
    \vskip\baselineskip
    \begin{subfigure}[b]{0.49\textwidth}
        \centering 
        \includegraphics[width=\textwidth]{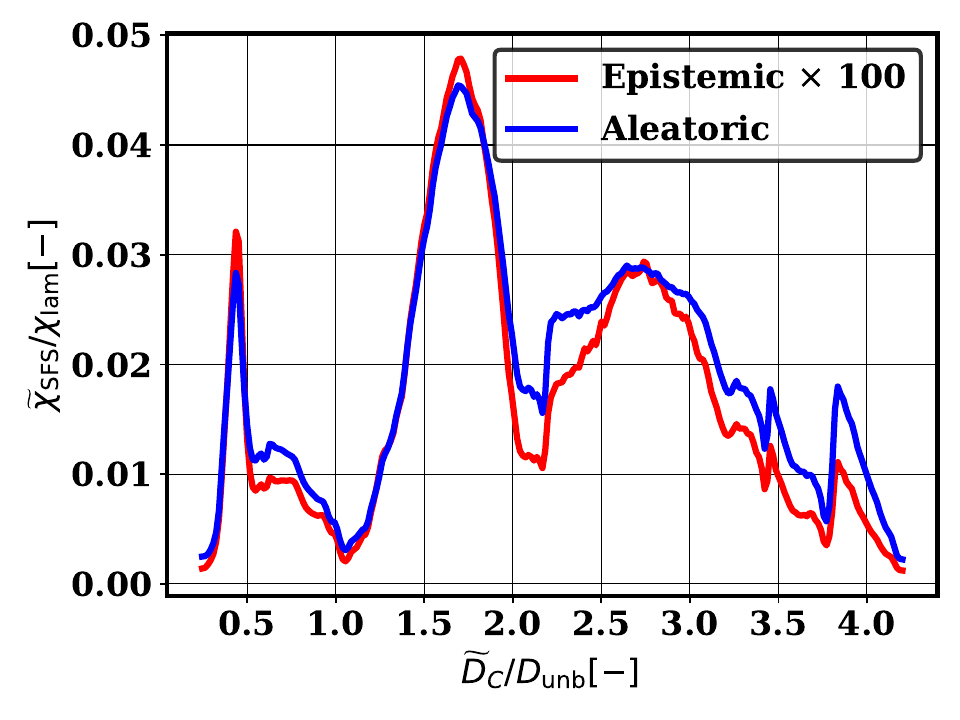}
        \caption{}
        \label{fig:epistemic-fd-isoprior}
    \end{subfigure}
    \caption{Conditional average of the aleatoric and epistemic uncertainty ($\times$ 100) with respect to (a) filtered progress variable, (b) filtered variance, and (c) filtered diffusion.}
    \label{fig:epistemic-stddev-isoprior}
\end{figure*}

To evaluate the impact of the prior selection as outlined in Sec.~\ref{sec:modeling-prior}, a model was trained using an isotropic Gaussian prior for the weights, i.e. $p(\mathbf{w})=\mathcal{N}(0,I)$. The conditional averages of the aleatoric and epistemic uncertainties for this model are presented in Figure~\ref{fig:epistemic-stddev-isoprior}. Compared to those presented in Fig.~\ref{fig:epistemic-stddev}, we observe that the resultant epistemic uncertainty is larger, which we attribute to the prior term's bias towards a model form that generates predictions with a mean of zero. There also appears to be a slight bias to the profile in Fig.~\ref{fig:epistemic-fc-isoprior} as compared to Fig.~\ref{fig:epistemic-fc}. Given the quantity of the data and the pre-processing steps to center the data, one might expect that an isotropic prior may be appropriate, and this indeed seems to be the case, however the results presented in the main text seem to have better agreement. As mentioned in Sec.~\ref{sec:modeling-prior}, specification of the prior is an ongoing area of research.

%% file: appendix-threshold.tex
\section{Effect of distance threshold on out of distribution query detection}
\label{sec:appendix-threshold}

In Algo.~\ref{alg:OODdetec}, a key variable for the \gls{OOD} of query points is the threshold $T$ which discriminates between query points in the \gls{OOD} regime and in-distribution regime. The value of $T$ depends on the magnitude of the distance metric $d_{\rm OOD}$, which itself depends on the magnitude of $\mu_{\rm OOD}$ and $\sigma_{\rm OOD}$. In Sec.~\ref{sec:OODquery}, the value of $T$ was therefore defined as $\alpha \sqrt{\mu_{\rm OOD}^2 + \sigma_{\rm OOD}^2}$, where it was chosen $\alpha = 0.6$. To define a reasonable value of $\alpha$, a sensitivity analysis was performed: the value of $\alpha$ was varied until the fields of in-distribution index became insensitive to $\alpha$.

\begin{figure*}[!htb]
    \centering
    \begin{subfigure}[b]{0.49\textwidth}
        \centering
        \includegraphics[width=\textwidth]{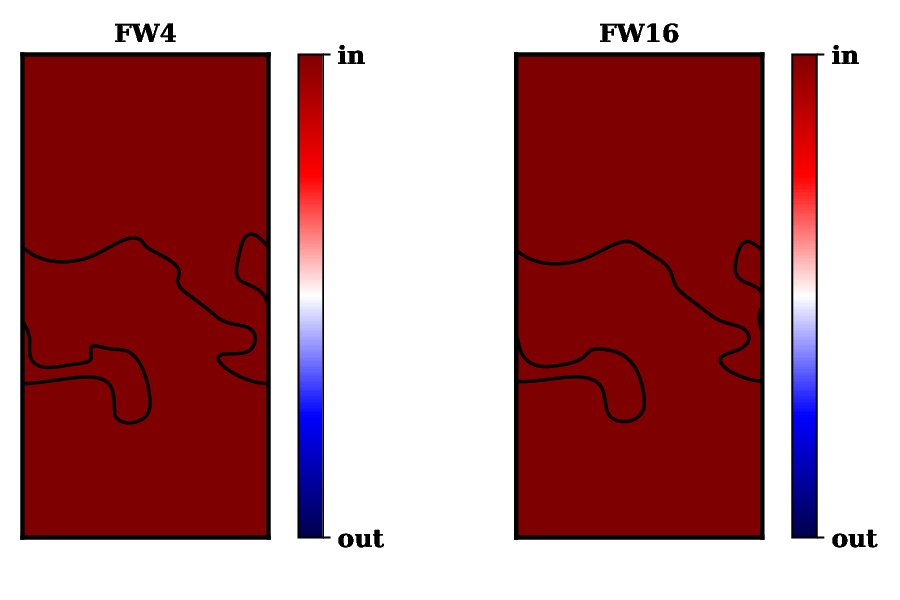}
        \includegraphics[width=\textwidth]{indist_BLE_1_5.eps}
        \includegraphics[width=\textwidth]{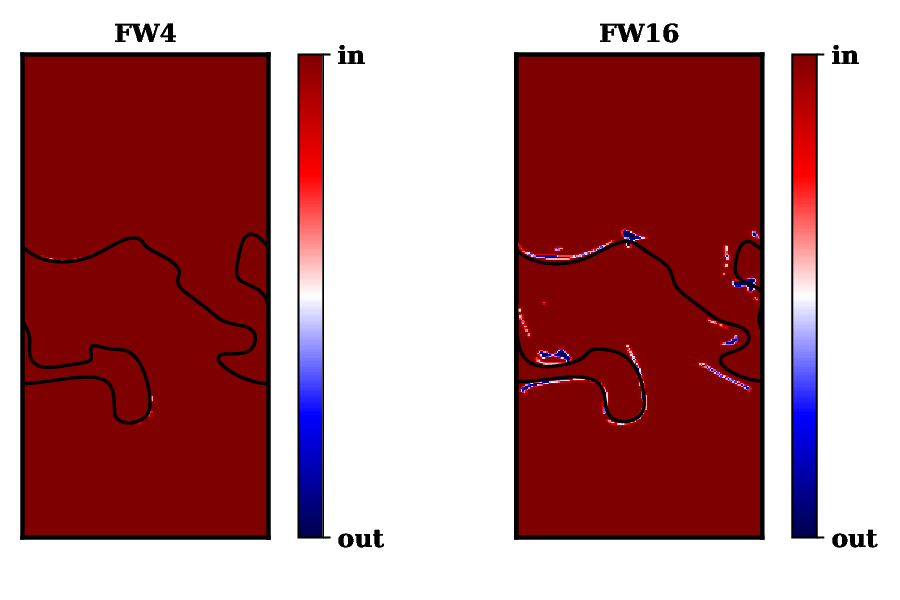}
        \caption{$B_{\rm Le}$ case.}
        \label{fig:BLE_thre}
    \end{subfigure}
    \hfill
    \begin{subfigure}[b]{0.49\textwidth}
        \centering
        \includegraphics[width=\textwidth]{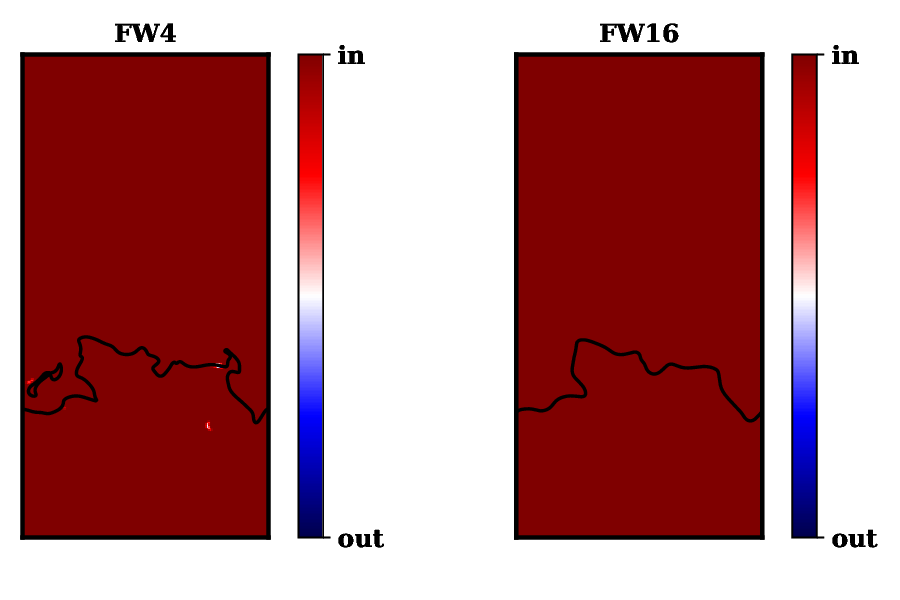}
        \includegraphics[width=\textwidth]{indist_DLE_1_5.eps}
        \includegraphics[width=\textwidth]{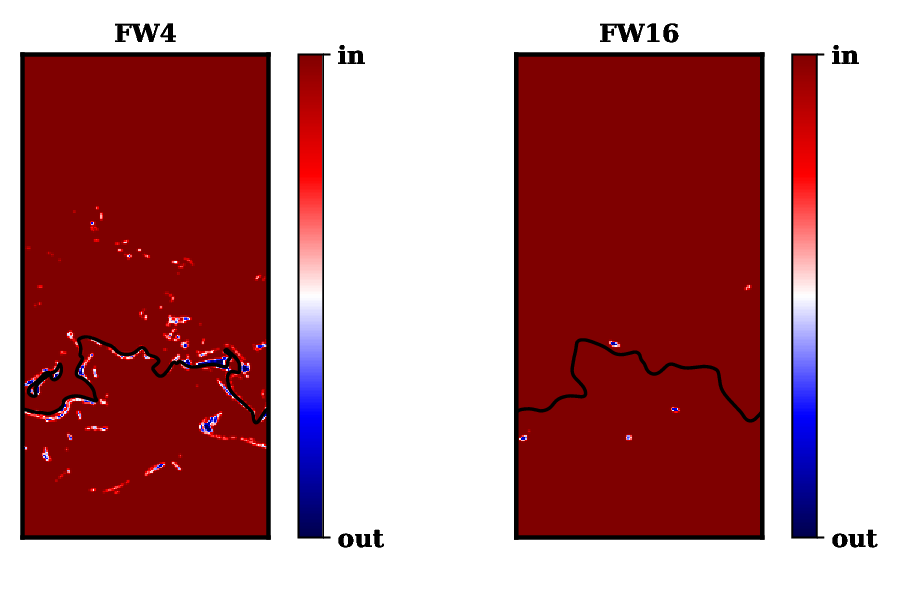}
        \caption{$D_{\rm Le}$ case.}
        \label{fig:DLE_thre}
    \end{subfigure}
    \caption{In-distribution index predicted by Algo.~\ref{alg:OODdetec} for filterwidth of 4 and 16. Top: $\alpha = 0.4$. Middle: $\alpha=0.6$. Bottom: $\alpha=0.8$. The isocontour denotes the flame contour based on the progress variable value.}
    \label{fig:OODthre}
\end{figure*}

Figure~\ref{fig:OODthre} shows that the value of $\alpha$ affects the in-distribution index fields. However, the in-distribution index fields is significantly less sensitive to $\alpha$ if $\alpha<0.6$ than if $\alpha>0.6$, which explains the choice adopted for $\alpha$ in Sec.~\ref{sec:OODquery}. For all values of $\alpha$, OOD queries detected are always clustered near the stretched portions of the flame front, which is indicative that the OOD queries coincide with inputs likely unseen in the training data. To systematically decide on the $\alpha$ threshold, one could first define OOD data based on the nearest neighbor metric and choose the threshold such that the probability of a false positive is lower than a prescribed amount.